\documentclass[reprint,twocolumn,amsmath,amssymb,floatfix]{revtex4-1}

\usepackage{graphicx}
\usepackage{hyperref}
\hypersetup{
     colorlinks=true,
     linkcolor=blue 
} 

\usepackage[T1]{fontenc} 
\usepackage[utf8]{inputenc}
\usepackage{lmodern}

\usepackage{subcaption}
\usepackage{float}
\usepackage{booktabs}
\usepackage{etoolbox}
\usepackage[separate-uncertainty=true]{siunitx}
\DeclareSIUnit\Molar{\textsc{m}}
\DeclareSIUnit{\mJ}{mJ}
\usepackage{upgreek}
\usepackage{chemmacros}
\chemsetup{
 formula = mhchem,
 greek = chemgreek,
 modules = thermodynamics,
 modules = redox,
 modules = reactions
}

\usepackage[symbol]{footmisc}
\renewrobustcmd{\bfseries}{\fontseries{b}\selectfont}
\renewrobustcmd{\boldmath}{}

\newcommand*{\etal}{et~al.}

\begin{document}
\author{Hanna Oher}
\affiliation{DEN-Service d’Etudes Analytiques et de Réactivité des Surfaces (SEARS), CEA, Université Paris-Saclay, F-91191 Gif-sur-Yvette, France}
\altaffiliation{Univ. Lille, CNRS, UMR 8523 - PhLAM - Physique des Lasers Atomes et Molécules, F-59000 Lille, France}
\author{Florent Réal}
\affiliation{Univ. Lille, CNRS, UMR 8523 - PhLAM - Physique des Lasers Atomes et Molécules, F-59000 Lille, France}
\author{Thomas Vercouter}
\affiliation{DEN-Service d’Etudes Analytiques et de Réactivité des Surfaces (SEARS), CEA, Université Paris-Saclay, F-91191 Gif-sur-Yvette, France}
\email{thomas.vercouter@cea.fr}
\author{Valérie Vallet}
\affiliation{Univ. Lille, CNRS, UMR 8523 - PhLAM - Physique des Lasers Atomes et Molécules, F-59000 Lille, France}
\email{valerie.vallet@univ-lille.fr}

\title[An \textsf{achemso} demo]
{Investigation of the luminescence of \ce{[UO2X4]^{2-}} (X=Cl, Br) complexes in organic phase using time-resolved laser-induced fluorescence spectroscopy and quantum chemical simulations}

\begin{abstract}
The luminescence properties of the \ce{[UO2Cl4]^{2-}} complex in an organic phase, especially the influence of large organic counter cations, have been studied by time-resolved laser-induced fluorescence spectroscopy (TRLFS) and ab initio modeling. The experimental spectrum was assigned by vibronic Franck-Condon calculations on quantum chemical methods on the basis of a combination of relativistic density functional approaches. The shape of the luminescence spectrum of the uranyl tetrachloride complex is determined by symmetrical vibrations and geometrical change upon emission. The possible change in the luminescence properties depending on the first and second uranyl coordination spheres was predicted theoretically for the \ce{[UO2Br4]^{2-}} and \ce{[R4N]2[UO2Cl4]} (\ce{[R4N]} = \ce{[Bu4N]}, \ce{[{A336}]}) systems. The computations reveal that for U(VI), the second coordination sphere has little influence on the spectrum shape, making speciation of uranyl complexes with identical first coordination-sphere ligands tedious to discriminate. The computed structural changes agreed well with experimental trends; theoretical spectra and peak attributions are in a good accordance with TRLFS and magnetic circular dichroism (MCD) data respectively.
\end{abstract}
\maketitle

\section{Introduction}
The knowledge of the stoichiometries and stabilities of the chemical species of uranium is of key importance for an understanding of the chemical reactivity of uranium in environmental or industrial situations. The speciation of uranium in solution has thus been a major topic of investigations by both experimental and theoretical methods. Among the powerful spectroscopic techniques, time-resolved laser-induced luminescence spectroscopy (TRLFS) has been widely used for the characterization of uranyl species in solutions and in solid phases, because of its high sensitivity to changes in the first coordination sphere of \ce{UO2^{2+}}. The luminescence spectra of uranyl complexes in solution show in general a narrow energetic range of about \SI{6000}{\per\cm}.  In this region only a single electronic transition between the initial and final states can be identified and it is vibrationally resolved with the bands corresponding to different vibrational quantum numbers~\cite{visnak2016uranium6}. In the case of mixtures of uranyl species, the interpretation of TRLFS data can be difficult as one needs to deconvolute overlapping emission spectra in similar time frames. Therefore, spectroscopic considerations are often insufficient to unambiguously determine the nature of the complexes. This is why a theoretical support based on quantum chemical modeling appears as a way to better validate TRLFS data interpretations by decomposing the different effects that might induce changes in the emission spectra of uranyl complexes.

Several research teams have tried to simulate emission spectra of uranyl complexes with quantum chemical methods~\cite{doi:10.1021/ct200419x, doi:10.1021/ic5006852, su2011uranyl}, on the basis of either wave-function theory (WFT) or density functional theory (DFT) and by taking into account relativistic effects~\cite{relat-Pyykko-CR1988-88-563,relat-Dyall-Book2007,relat-Autschbach-JCP2012-136-150902,relat-Reiher-Book2014} (scalar relativistic effects and spin-orbit coupling), as they are important for actinide complexes. While WFT methods are well-suited for accurate calculations, the high computing costs limit the simulations to small clusters of heavy elements such as uranium. DFT approaches are more appropriate for cost-effective calculations applied to actinides but can be challenging with regard to their accuracy for the description of the electronic ground and excited states and their ability to reproduce low-lying electronic transitions and their vibronic resolutions. Tecmer~{\etal}~\cite{tecmer2011electronic} have evaluated the accuracy of pure and hybrid exchange-correlation functionals for the bare uranyl cation isoelectronic uranium triatomics, arguing that hybrid functionals can be used for quantitative prediction of the low-lying excited states of uranyl. Uranyl tetrachloride complexes have been thoroughly studied as a reference system because direct comparisons with luminescence data from crystals are possible~\cite{watkin1991structure}. Moreover, this complex can also be stabilized in nonaqueous solvents~\cite{B317002K, doi:10.1021/ic701752j, sornein2006uranyl} and is of interest in some solvent extraction protocols~\cite{reillerA336termodynamics,helle2015liquid}. The structure of the uranyl tetrachloride is well-established, with four chloride ions in the equatorial plane of the \ce{UO2^{2+}} moiety, resulting in a $D_{4h}$ symmetry. The effects of second-sphere countercations, solvent molecules, or solvating agents also have to be considered, in order to quantify their influence on the structures, the electronic state energies, and the vibronic progressions. 

In this work we compare the results of \textit{ab initio} calculations to experimental data obtained from a $n$-dodecane solution in which the uranyl tetrachloride complex was extracted by a tetraalkylammonium molecule (Aliquate \textsuperscript{\textregistered} 336). The methodology is a step-by-step approach in order to validate the hypotheses and approximations we made. The computations of the ground and the excited state structural and vibrational parameters are necessary for an understanding of the theoretical and experimental spectra. Because the main effect on the luminescence data is commonly accepted to originate from the first coordination sphere, the chloride ligands were substituted by bromides to discuss the trends, validate our method, and evaluate its applicability to other types of uranyl complexes. We will also quantify the effects of outer-sphere counterions and long-range solvation effects on the computed spectra.

\section{Experimental and Computational Details}
\subsection{Sample Preparation: \ce{[A{336}]_2[UO_2Cl_4]} in $n$-Dodecane}
The sample preparation method has been adapted from Hellé~{\etal}~\cite{helle2014liquid, helle2015liquid}. Aliquate \textsuperscript{\textregistered} 336 (\SI{98}{\percent}) was purchased from Alfa Aesar. HCl (\SI{32}{\percent}), 1-decanol (\SI{99}{\percent}) and \textit{n}-dodecane were purchased from Sigma-Aldrich. \ce{HClO4} was purchased from Merck. All reagents were used as received without further purification. 
A stock solution of uranium(VI) was prepared by dissolution of \ce{U3O8} in a hot perchloric acid solution. The U(VI) concentration was checked by inductively coupled plasma mass spectrometry (ICP-MS). The aqueous solution was prepared by dilution of this stock solution into a \SI{5}{\Molar} hydrochloric acid solution to get a uranium (VI) concentration of \SI{e-5}{\Molar}. Deionized water (Alpha-Q, Millipore, \SI{18.2}{\mega\Omega\cm}) was used for the preparation of all aqueous solutions. 

The organic solution was prepared by dissolving weighed amounts of Aliquate \textsuperscript{\textregistered} 336  and 1-decanol in sufficient amount of \textit{n}-dodecane to reach a concentration of \SI{e-2}{\Molar} of Aliquate \textsuperscript{\textregistered} 336 with \SI{1}{\percent} of 1-decanol. It was then pre-equilibrated by contact with a uranium-free \SI{5}{\Molar} hydrochloric acid solution during \SI{2}{\hour} of shaking and separated. 

For uranium extraction, \SI{2}{\mL} of the aqueous solution was contacted with an equal volume of the pre-equilibrated organic solution. The mixture was shaken in a thermomixer at \SI{20}{\degreeCelsius} during \SI{1}{\hour}, and about \SI{2}{\mL} of the organic phase was sampled after \SI{2}{\hour} of decantation for the spectroscopic measurements. 

\subsection{Time-Resolved Laser-Induced Fluorescence Spectroscopy}
The sample was put in a \SI{1}{\cm} path length quartz cuvette that was placed in a TRLFS setup as described afterward. The excitation wavelength was provided by a tunable OPO system (PantherEx OPO, Excel Technology) pumped by a Nd/YAG laser at \SI{355}{\nm} (Surelite-I, Excel Technology). The excitation was tuned to $\lambda_{ex}$ = \SI{427}{\nm} which corresponds to a maximum of absorption by uranium(VI) in our samples. The \SI{5}{\ns} laser pulses were generated at \SI{10}{\hertz} for an energy of about \SI{3.15}{\mJ}. The detection setup has already been described elsewhere ~\cite{doi:10.1021/ic701379q}. The luminescence signal was collected during a gate width of \SI{200}{\us}, with a gate delay of \SI{100}{\ns} after the excitation by the laser pulse, the delay value being carefully chosen as discussed in the Supporting Information. The luminescence spectrum of the sample was recorded at room temperature (\SI{22\pm1}{\degreeCelsius}) from the accumulation of \num{1000} scans. The background noise was subtracted by the software from the recorded spectrum of a \ce{[A{336}]_2[UO_2Cl_4]} in $n$-dodecane sample. 

\subsection{Computational Details}

Since our aim is to use quantum chemical methods to elucidate the luminescence band shapes of the complexes, several data need to be calculated. The ground-state and first-excited-state geometries, their associated harmonic frequency spectra, and Hessian matrices have to be computed in order to derive the overlap integrals between the vibrational wave-functions associated with the ground and excited states - Franck-Condon factors (FCFs). 
The FCFs were computed using the ezSpectrum~3.0~\cite{ezSpectrum} program by taking all necessary data generated by the ab initio packages described below. The Duschinsky rotations were used as implemented in the program. The numbers of vibrational quanta in excited and ground state were selected to be one and five, respectively. All of the spectra were computed at \SI{300}{\kelvin}. For the larger systems, all normal modes with vibrational frequencies larger than \SI{1000}{\per\cm} were excluded from the FCF calculations, to keep the computational costs affordable.

\begin{figure}[!h]
	\includegraphics[width=\linewidth]{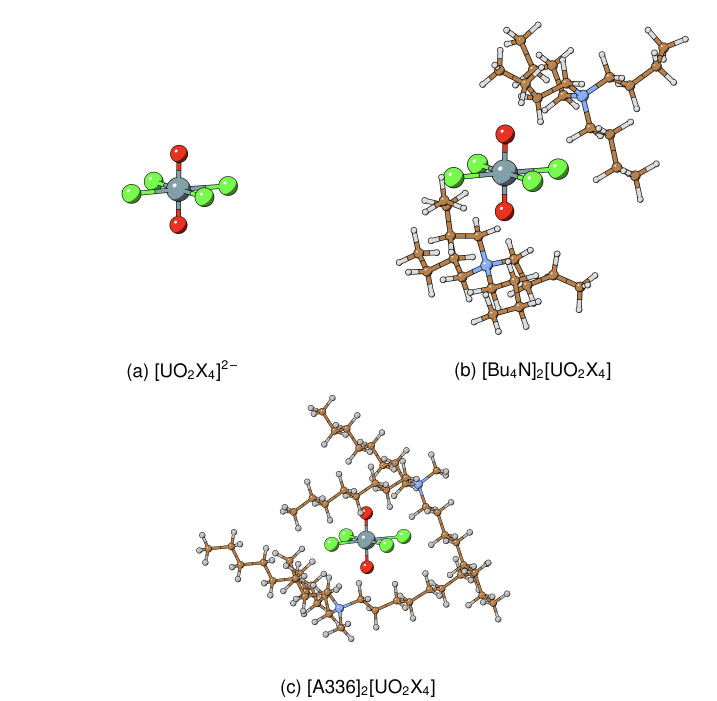}
	\caption{Structures of \ce{[R4N]2[UO2X4]} complexes (\ce{[R4N]} = \ce{[Bu4N]}, \ce{[{A336}]}; \ce{X}=\ce{Cl}, \ce{Br}) in the gas phase optimized at the DFT/PBE0 level of theory.}
	\label{fig:structures}
\end{figure}

\paragraph{Model Systems.} 
The interactions between uranyl and its first and second coordination spheres might affect the electronic structure of the uranyl unit. As the influence of the chloride ligands in the\ce{UO2Cl4^{2-}} complex was excellently reviewed by different experimental~\cite{doi:10.1021/jp071061n, doi:10.1080/00268978100102391, doi:10.1021/ic701752j, hopkins2001spectroscopy, B317002K} and theoretical methods~\cite{An-Ruiperez-JPC2010-114-3615-3621, doi:10.1063/1.2735297, doi:10.1063/1.2121608, doi:10.1021/jp003032h, gomes2013towards}, this system was selected as a benchmark to quantify the effect of ligands in the first coordination sphere by substituting chlorides with bromides and the effect of the counterions in the second coordination sphere: i.e., the quaternary ammonium cations. Furthermore, to discuss the importance of long-range solvent effects, the model systems were computed in the gas phase and with inclusion of solvent effects (\textit{n}-dodecane and acetone). The structures are represented in Fig.~\ref{fig:structures}

\paragraph{Structures of the Ground and Excited states and Harmonic Frequencies.}
For the sake of keeping the computational costs within a scale of 1 week and of simplifying the data analysis, the structures of the uranyl tetrahalide complexes were enforced to have $D_{4h}$ symmetry, whereas no symmetry constraints were applied to the complexes with a second coordination sphere. All of the ground-state molecular geometries were optimized including the relativistic effects at the spin-free level in the gas phase and with solvent effects using density functional theory (DFT). The Kohn-Sham equation was solved using the hybrid PBE0  functional~\cite{pbe0}. 
The structure of the first low-lying excited state was optimized using the time-dependent (TD)-DFT/PBE0 method as implemented in the Turbomole V7.3 2018~\cite{TURBOMOLE}  and Gaussian 16~\cite{g16} codes. The vibrational harmonic frequencies were computed using either an analytic Hessian matrix or numerical finite differences of the gradient. All geometries considered for the vibronic spectra calculations represent true minima as they have no imaginary frequencies. The optimized structures are available in the Supporting Information.

In these calculations, def2-TZVP (second generation of triple-$\zeta$ polarization quality) Karlsruhe basis sets~\cite{weigend1998ri-basissetOCNH, weigend2005balanced-Clbasisset} have been used for all light elements (H, C, N, O, Cl). For the heaviest atoms, small-core relativistic effective core pseudopotentials (RECP) were used: namely, the 60-core-electron RECP for uranium~\cite{kuchle1994ECP1, cao2003ECP2} along with the def-TZVP (first generation of triple-$\zeta$ polarization quality) basis set~\cite{eichkorn1997def-TZVP}, and the  28-core electrons RECP for bromine with the associated aug-cc-pVTZ-PP basis set~\cite{aug-cc-pVTZ-PP}. To speed up the calculations, the resolution of the identity approximation to compute the Coulomb integrals (RI-J)~\cite{ deglmann2004nuclear, bauernschmitt1997calculation} with appropriate auxiliary basis sets~\cite{weigend2006accurate, eichkorn1997def-TZVP} was employed. 

\paragraph{Vertical Absorption and Emission Energies.}
On the basis of the benchmark calculations carried out by Tecmer~{\etal}~\cite{tecmer2011electronic, doi:10.1021/jp3011266, tecmer2013reliable} on a series of uranium(VI)-based compounds, the CAM-B3LYP~\cite{yanai2004new} exchange-correlation functional was found to be more accurate than the PBE0 functional for uranyl valence transition energies. Thus, to accurately position the ``hot bands'', the vertical excitation and emission energies with and without accounting for solvent effects were obtained from the ground- and excited-state structures, respectively, with TD-DFT single-point calculations with the latter functional and the Amsterdam Density Functional package (ADF2018.01)~\cite{ADF2018authors}. All atoms were described by TZ2P Slater-type basis sets~\cite{basis-Van-Lenthe-JCC2003-24-1142} (triple-$\zeta$ with two polarization functions quality), without the frozen core approximation. The scalar relativistic (SR) and spin-orbit coupling (SOC) effects were accounted for by the ZORA Hamiltonian~\cite{lenthe1993relativistic}.

\paragraph{Continuum Solvent Models.}
Long-range solvent effects were modeled by polarized continuum medium models, with two similar flavors: the continuum polarizable conductor model (CPCM)~\cite{Barone:1998aa-cpcm, doi:10.1002/jcc.10189} as implemented in the Gaussian 16 program and the conductor-like screening model (COSMO) model~\cite{COSMO-ADF, COSMO_ADF_1, COSMO-ADF_2} as implemented in the ADF package. The relative permittivity values ($\varepsilon_r$) used for $n$-dodecane and acetone are 2.006 and 20.493, respectively.

\section{Results and Discussions}

\subsection{Experimental Luminescence Spectra}
\begin{figure}[!htbp]
	\centering
	\includegraphics[width=\linewidth]{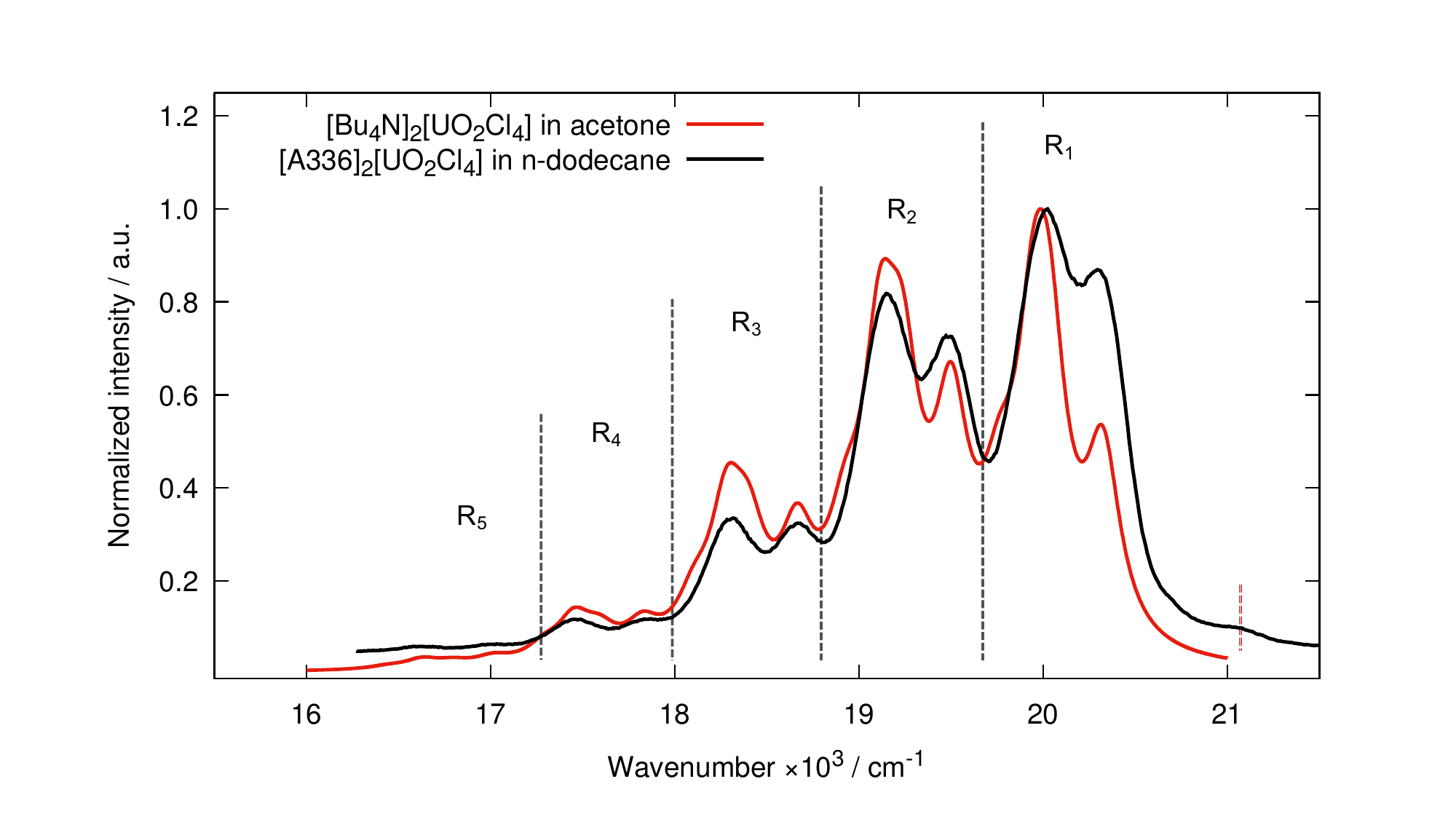}
	\caption{Recorded time-resolved luminescence spectra of \ce{[Bu4N]2[UO2Cl4]} in acetone (the maxima taken from G\"{o}rller-Walrand ~{\etal}~[\citenum{B317002K}] and the Lorentzian shape computed) and \ce{[{A336}]2[UO2Cl4]} in \textit{n}-dodecane (this work). The vertical dashed red line shows the ``hot band'' position and $R_n, (n=1-5)$ corresponds to the vibronic progression region. Details on the spectral data are available in Table~S1 of the Supporting Information.}
	\label{fig:experimental_trlfs}
\end{figure}

The time-resolved luminescence spectrum of the uranyl sample in the presence of chloride ions and Aliquate 336 in \textit{n}-dodecane with \SI{1}{\percent} of 1-decanol is shown on Figure~\ref{fig:experimental_trlfs} (black line). It is superimposed with the spectrum (red line) acquired by G\"{o}rller-Walrand~{\etal}~\cite{B317002K} under similar conditions and in acetone. The spectrum also compares well with those obtained earlier in chloroaluminate~\cite{hopkins2001spectroscopy}, tetraalkylammonium~\cite{sornein2006uranyl} and pyrrolidinium~\cite{doi:10.1021/ic701752j} ionic liquids.
The spectrum of \ce{[{A336}]2[UO2Cl4]} in \textit{n}-dodecane in Figure~\ref{fig:experimental_trlfs} shows an electronic transition  ("hot band'') at low energy about \SI{21000}{\per\cm} followed by a series of vibronic peaks in the \SIrange{20300}{16000}{\per\cm} range, which is typical of the \ce{[UO2Cl4]^{2-}} species with a $D_{4h}$ coordination symmetry~\cite{denning1976electronic}. A monoexponential decay with a fluorescence lifetime of \SI{0.3}{\us} was measured for the uranium(VI) sample in a presence of chloride ions, Aliquate \textsuperscript{\textregistered} 336 and \textit{n}-dodecane. This confirms the formation of a unique complex, which was assumed to be \ce{[{A336}]2[UO2Cl4]} with four chloride ions coordinated to uranyl in its equatorial plane, in line with the extracted complex stoichiometry~\cite{reillerA336termodynamics}. 

The decomposition of the \ce{[{A336}]2[UO2Cl4]} luminescence spectra has been performed by Reiller~{\etal}~\cite{reillerA336termodynamics}. The resolution of the presently used apparatus did not allow good resolution of some of the transitions and revealed only the wide components which cover several vibronic progressions in one peak. Moreover, a difference in the relative intensities is observed at around \SI{20295}{\per\cm} in $n$-dodecane due to the long gate width selected to collect more luminescence signals in our measurements. The spectrum in $n$-dodecane is shifted to the blue side by \SI{30}{\per\cm}, in comparison to that in acetone, and only by \SI{5}{\per\cm} with respect to that in \ce{[Bu3MeN][Tf2N]} ionic liquid~\cite{sornein2006uranyl}. Thus, from the measurements, one can conclude that neither the counterion nor the type of solvent has any significant influence on the position of maxima of the luminescence bands. The change in the relative intensities of the transitions along the vibronic progression has been attributed by Sornein~{\etal}~\cite{sornein2006uranyl} to the formation of \ce{C\bond{-}H\bond{...}Cl} hydrogen bonding between the chloride of the \ce{[UO2Cl4]^{2-}} moiety and a hydrogen atom of a cation present in the ionic liquid. As we are using an aprotic solvent (\textit{n}-dodecane), this effect may also be present in our uranyl tetrachloride sample.

The spacing between the vibronic progressions of the luminescence spectrum corresponds to the ground-state uranyl stretching frequency value $\nu_{s}$. One can extract the $\nu_{s}$ value by only taking into account the spacing between the vibrational maxima of the same nature for \ce{[{A336}]2[UO2Cl4]} in $n$-dodecane. The resulting $\nu_{s}$ value amounts to \SI{836\pm18}{\per\cm}, which is in good agreement with the values, \num{823\pm14}, \num{825}, and \SI{850\pm25}{\per\cm} observed in acetone~\cite{B317002K}, \ce{[C_4mim][Tf_2N]}~\cite{doi:10.1021/ic701752j}, and in  \ce{[Bu3MeN][Tf2N]}~\cite{sornein2006uranyl}, respectively. This vibration corresponds to a Raman-active vibration of the uranyl unit and can be correlated to the \ce{U-O_{yl}} bond length. Using the empiric relation of Bartlett and Cooney~\cite{bartlett1989determination}
\begin{equation}
R_{\ce{U-O_{yl}}} /(\mathbf{\AA})=106.5 \times \nu_{s}^{-2/3}+0.575
\end{equation}
we calculated the \ce{U-O_{yl}} bond length to be equal to \SI{1.77\pm0.01}{\angstrom} (using $\nu_{s}$ = \SI{836\pm18}{\per\cm}) for our uranyl sample. This value agrees with all experimental data given in Table~\ref{table:structures_tetrahalides} and can therefore serve as reference data to assess the accuracy of the \textit{ab initio} calculations we will now discuss.

\subsection{Ground- and Excited-State Structures of the Uranyl Tetrahalide Complexes}
\paragraph{Uranyl Tetrahalide Dianions.}
The structure of the uranyl tetrachloride dianion has been well-studied by the variety of theoretical and experimental techniques~\cite{dau2013, doi:10.1021/jp3011266, doi:10.1063/1.2121608, watkin1991structure, doi:10.1021/ic051098r, wilson2011structural, doi:10.1021/jp071061n, deifel2010supramolecular}. The inclusion of all effects to mimic the experimental conditions is challenging for \textit{ab initio} calculations mainly because of the difficulties in building chemically relevant chemical models and the fast growth of the computational costs as the system size grows. The present quantum chemical study was performed on models with increasing chemical complexity to approach the solution sample. 

The spin-free optimized ground-state and first excited-state distances between uranium and the coordinated atoms of the \ce{[UO2Cl4]^{2-}} and \ce{[UO2Br4]^{2-}} together with experimental data are shown in Table~\ref{table:structures_tetrahalides}. 

In the \ce{[UO2Cl4]^{2-}} gas-phase PBE0 calculations, the ground state  \ce{U-O_{yl}} bond length of \SI{1.758}{\angstrom} is in a good agreement with \SI{1.766}{\angstrom} obtained with the CCSD(T) wave-function method, while the \ce{U-Cl} bond length of \SI{2.714}{\angstrom} is shorter than \SI{2.735}{\angstrom}. Moreover, the PBE0 geometries are in good accordance with the all-electron CAM-B3LYP results reported by Tecmer~{\etal}~\cite{doi:10.1021/jp3011266}. This gives us confidence in the ability of PBE0 to provide fairly accurate geometries of the uranium(VI)-containing complexes. The inclusion of solvent effects on \ce{[UO2Cl4]^{2-}}  does not significantly change the  \ce{U-O_{yl}} bond distance in comparison to the gas-phase calculations, but it shortens the \ce{U-Cl} bond length as the solvent polarity increases. Keeping in mind that we do not have strictly the same conditions as in the experiment, it is worth examining trends in both theoretical and experimental data. A comparison of the experimental crystal structure of \ce{Cs_2UO_2Cl_4}~\cite{watkin1991structure} with the structure of uranyl tetrachloride dianion in acetonitrile~\cite{doi:10.1021/ic051098r} shows that the  \ce{U-O_{yl}} bond length is almost the same under both conditions, while \ce{U-Cl} values are found to be longer in solution by about \SI{0.01}{\angstrom}. 

To quantify the influence of the ligand nature in the first coordination sphere, the chloride ligands were replaced by bromide ligands, which have the same type of bonding with uranium but larger ionic radii. The uranyl ion coordinated by four bromide ligands has been prepared in crystal form~\cite{wilson2011structural}, while it is hardly stabilized in solution. The experimental structure of the \ce{Cs2UO2Br4} crystal shows that the \ce{U-O_{yl}} bond distance (\SI{1.777}{\angstrom}) is almost equal to that found in the chloride homologue \ce{Cs2UO2Cl4} ($\mathrm{R_{U-O_{yl}}}$ = \SI{1.774}{\angstrom}), whereas the \ce{U-Br} distance is longer by \SI{0.149}{\angstrom} than that of \ce{U-Cl}. The gas-phase geometries of \ce{[UO2Cl4]^{2-}} and \ce{[UO2Br4]^{2-}} follow this trend; the substitution of chloride by bromide ligands leads to an insignificant stabilization of the \ce{U-O_{yl}} bond, and the \ce{U-Br} bond distance is longer by \SI{0.169}{\angstrom} than that of \ce{U-Cl}, as the \ce{U-Br} bond is weaker than the \ce{U-Cl} bond. 
\begin{table*}[htbp!]
		\centering
 	\caption{Ground- and Excited State Geometries of \ce{[UO2X4]^{2-}} (\ce{X} = \ce{Cl}, \ce{Br}) Compared to Selected Previous Results}
 		\label{table:structures_tetrahalides}
	\begin{tabular}{llllll}
		\toprule 
		& $\mathrm{R_{U-O_{yl}}}$, \r{A}  & $\mathrm{R_{U-X}}$, \r{A}   & Compound/media           & Method          & Ref.                                     \\ \hline
		\multicolumn{6}{l}{Ground state}                                                                                      \\ \hline
		\ce{[UO2Cl4]^{2-}} & 1.758 & 2.714 & gas-phase       & R-ECP/PBE0  & this work                                     \\
		& 1.766 & 2.735 & gas-phase        & CCSD(T)         & [\citenum{dau2013}]              \\
		& 1.764 & 2.712 & gas-phase       & all-electron/CAM-B3LYP       & [\citenum{doi:10.1021/jp3011266}] \\
		& 1.774 & 2.671 & \ce{Cs2UO2Cl4}       & X-ray           & [\citenum{watkin1991structure}]   \\
		& 1.759 & 2.699 & CPCM n-dodecane & R-ECP/PBE0      & this work                                     \\
		& 1.761 & 2.683 & CPCM acetone    & R-ECP/PBE0      & this work                                     \\
		& 1.770  & 2.680  & acetonitrile    & EXAFS           & [\citenum{doi:10.1021/ic051098r}] \\
		&       &       &                 &                 &                                               \\
		\ce{[UO2Br4]^{2-}} & 1.750 & 2.883 & gas-phase       & R-ECP/PBE0      & this work                                     \\
		& 1.749 & 2.905 & gas-phase       & R-ECP/CAM-B3LYP & this work                                     \\
		& 1.777 & 2.820  & \ce{Cs2UO2Br4}       & X-ray           & [\citenum{wilson2011structural}]  \\ \hline
		\multicolumn{6}{l}{Excited state}                                                                                     \\ \hline
		\ce{[UO2Cl4]^{2-}} & 1.789 & 2.721 & gas-phase       & R-ECP/PBE0      & this work                                     \\
		& 1.844 & 2.685 &  \ce{Cs2UO2Cl4} & Two-photon absorption  & [\citenum{doi:10.1021/jp071061n}] \\
		& 1.790  & 2.706 & CPCM n-dodecane & R-ECP/PBE0      & this work                                     \\
		& 1.791 & 2.692 & CPCM acetone    & R-ECP/PBE0      & this work                                     \\
		&       &       &                 &                 &                                               \\
		\ce{[UO2Br4]^{2-}} & 1.778 & 2.895 & gas-phase       & R-ECP/PBE0      & this work                                     \\
		& 1.780  & 2.918 & gas-phase       & R-ECP/CAM-B3LYP & this work        \\  \bottomrule             
	\end{tabular}
\end{table*}

The lowest triplet excited state in \ce{[UO2X4]^{2-}} complexes corresponds to an excitation of an electron out of an orbital that is a mixture of the bonding $\sigma_u$ orbital of the uranyl unit and the halide valence $p$ orbitals into the nonbonding $\delta_u$ uranium orbital. When the transition occurs the \ce{U-O_{yl}} bond weakens and the excited state potential curve shifts along the symmetric \ce{U-O_{yl}} mode to greater \ce{U-O_{yl}} distances and becomes flatter~\cite{doi:10.1063/1.2735297}. Looking at the excited-state geometries of the uranyl tetrahalide dianions, one can note a lengthening of the \ce{U-O_{yl}} bond by \num{0.031} and \SI{0.028}{\angstrom} in \ce{[UO2Cl4]^{2-}} and \ce{[UO2Br4]^{2-}}, respectively, since the excitation depopulates the \ce{U-O_{yl}} bonding orbital. The differences in the \ce{U-X}, (X = Cl,Br) bonds are small, about \SI{0.010}{\angstrom} on average. From highly resolved low-temperature two-photon absorption spectra of \ce{Cs2UO2Cl4}, Denning~{\etal}~\cite{denning1976electronic,doi:10.1021/jp071061n} determined that the \ce{U-O_{yl}} and \ce{U-Cl} bonds elongate by \num{0.070} and \SI{0.014}{\angstrom}, respectively. Inclusion of long-range solvent effects affect the excited state structure in the same way as for the ground state. 

The analysis of the vibrational frequencies is very important for the characterization of the theoretical luminescence spectra, as specific vibrations appear in the vibronic progression. Some of the ground-state vibrational frequencies are responsible for the band spacing in the experimental luminescence spectrum, while the atomic displacement between the ground and excited-state geometries is responsible for the intensity ratio. The \ce{U-X} ($\nu_\mathrm{U-X}$) and  \ce{U-O_{yl}}  ($\nu_\mathrm{s}$) symmetrical stretching modes and \ce{O_{yl}-U-O_{yl}} bending mode ($\nu_\mathrm{b}$) contribute to the luminescence spectral shape to a large extent, and their values for the \ce{[UO2Cl4]^{2-}} and \ce{[UO2Br4]^{2-}} complexes are given in Table~\ref{table:freq}. The \ce{U-O_{yl}} asymmetrical stretching mode ($\nu_\mathrm{a}$), which is vibronically silent, is also shown to allow discussion of trends. Using the vibrational perturbation theory~\cite{spectro-Barone-JCP2004-120-3059,spectro-Barone-JCP2005-122-014108} as implemented in Gaussian~16~\cite{g16}, we have computed the anharmonic corrections, which turned out to be small: on the order of \num{5} and \SI{3}{\per\cm} for the ground and excited states, respectively (see Table~S2 in the Supporting Information).

For the gas-phase \ce{[UO2Cl4]^{2-}} model, the computed $\nu_\mathrm{U-Cl}$, $\nu_\mathrm{b}$, $\nu_\mathrm{s}$ and $\nu_\mathrm{a}$ frequencies are smaller in the excited state than in the ground state by 4, 4, 80, and \SI{115}{\per\cm}, respectively. The PBE0 calculations reproduce with an impressive accuracy the measured red shift of the stretching mode, $\nu_\mathrm{s}$, \SI{82}{\per\cm}, in \ce{Cs2UO2Cl4}, and underestimates that for the bending mode, \SI{20}{\per\cm}. The latter discrepancy has negligible impact as the $\nu_\mathrm{s}$ dominates the vibronic progressions visible in the luminescence spectra.

\begin{table*}[!htbp]
	\centering
	\caption{Ground- and Excited-State Vibrational Frequencies (in $\mathrm{cm^{-1}}$) of the \ce{[R4N]_2[UO_2X_4]}, (\ce{[R4N]} = \ce{[Bu4N]}, \ce{[{A336}]}) Compounds}
	\label{table:freq}
	\begin{tabular}{lllllll}
		\toprule
		System                         & $\nu_\mathrm{U-X}$                                         & $\nu_\mathrm{b}$ & $\nu_\mathrm{s}$ &$\nu_\mathrm{a}$ & Method                                        & Ref.                                              \\ \hline
		\multicolumn{7}{l}{Ground state}                                                                                                                                                                                        \\ \hline
		\ce{[UO_2Cl_4]^{2-}}     & 237                                               & 275       & 894      & 974      & R-ECP/PBE0 gas-phase                          & this work                                        \\
		    & 235                                               & 269       & 889      & 960      & R-ECP/PBE0 n-dodecane                          & this work                                        \\
		     & 241                                               & 265       & 889      & 949      & R-ECP/PBE0 acetone                         & this work                                        \\
		\ce{Cs2UO2Cl4}        & 264 [\citenum{doi:10.1021/jp071061n}] & 250       & 832      & 915      & Raman and IR in solid                         & [\citenum{denning1976electronic}]    \\
			\multicolumn{7}{l}{}                                                                                                                                                                                                    \\
		\ce{[Me4N]2[UO2Cl4]}     &                                                   &           & 831      & 909      & Raman and IR in solid                         & [\citenum{schnaars2013structural}]   \\
		\ce{[Et4N]2[UO2Cl4]}    & 240                                               & 263       & 869      &          & Raman in solid                                & [\citenum{newbery1969vibrational}]   \\ 
		\ce{[Bu4N]2[UO2Cl4]} & 270                                               & 286       & 869      & 945      & R-ECP/PBE0 gas-phase                          & this work                                        \\
		& 258                                               & 262       & 833      & 919      & Raman and IR in \ce{CH2Cl2} solution & [\citenum{gal1984mid}]               \\
		\ce{[{A336}]2[UO2Cl4]}  & 266                                               & 288       & 876      & 957      & R-ECP/PBE0 gas-phase                          & this work                                        \\ 
			\multicolumn{7}{l}{}                                                                                                                                                                                                    \\
			\ce{[UO2Br4]^{2-}} &144& 262&907&988& R-ECP/PBE0 gas-phase     & this work      \\ 
			\ce{[C7H16NO2][UO2Br4]}&173&252&826&904&Raman and IR in solid& [\citenum{marzotto1974preparation}] \\ 
			 \hline
		\multicolumn{7}{l}{Excited state}                                                                                                                                                                                       \\ \hline
		\ce{[UO2Cl4]^{2-}}     & 233                                               & 271       & 814      & 859      & R-ECP/PBE0 gas-phase                          & this work                                        \\
		     & 236                                               & 268       & 811      & 854      & R-ECP/PBE0 n-dodecane                          & this work                                        \\
		     & 240                                               & 263       & 808      & 848      & R-ECP/PBE0 acetone                         & this work                                        \\
		\ce{Cs2UO2Cl4}        &                                                   & 230       & 750      & 830      &N.A.                                & [\citenum{B317002K}]                 \\
			\multicolumn{7}{l}{}                                                                                                                                                                                                    \\
		\ce{[Bu4N]2[UO2Cl4]} & 266                                               & 281       & 799      & 845      & R-ECP/PBE0 gas-phase                          & this work                                        \\
		\ce{[{A336}]2[UO2Cl4]}  & 261                                               & 284       & 802      & 854      & R-ECP/PBE0 gas-phase                          & this work                               \\ 
		\multicolumn{7}{l}{}                                                                                                                                                                                                    \\
		\ce{[UO2Br4]^{2-}}  &151&264&876&935& R-ECP/PBE0 gas-phase &this work       \\ 
			\bottomrule        
	\end{tabular}
\end{table*}

Turning now to the vibrational frequencies obtained with solvent effects, we observed an opposite behavior of the \ce{U-Cl} and \ce{U-O_{yl}} frequencies. With an increase in polarity the $\nu_\mathrm{U-Cl}$ stretching frequencies insignificantly increase (4 and \SI{7}{\per\cm}) for the ground and excited states, respectively, while the uranyl $\nu_\mathrm{b}$, $\nu_\mathrm{s}$, and $\nu_\mathrm{a}$ frequencies decrease by \SIrange{5}{25}{\per\cm} in both states. This shift in the ground-state frequencies should be observable in the computed luminescence spectra, namely in the spacing between the vibrationally resolved peaks,  while the change in the excited-state frequencies should only affect the vibronic intensities. 

The substitution of chlorides by bromides in the uranyl equatorial plane lead to the following changes in the vibrational spectra. In comparison to uranyl tetrachloride, the $\nu_\mathrm{b}$ value decreased by less than \SI{10}{\per\cm} in both the ground and excited states. The $\nu_\mathrm{s}$ values increased by 13 and \SI{62}{\per\cm} in the ground and excited states, respectively. Similarly, the $\nu_\mathrm{a}$ values increased by \SI{14}{\per\cm} for the ground state and \SI{76}{\per\cm} for the first excited state. The \ce{U-Br} stretching mode is \SI{83}{\per\cm} lower in both the ground and excited states than the \ce{U-Cl} symmetric stretching vibration. This is in line with the ground-state experimental trends measured for \ce{Cs2UO2Cl4} and \ce{[C7H16NO2][UO2Br4]} (91, 2, 6 and 9 $\mathrm{cm^{-1}}$ for $\nu_\mathrm{U-Br}$, $\nu_\mathrm{b}$, $\nu_\mathrm{s}$, and $\nu_\mathrm{a}$, respectively). These changes indicate that the bromide ligand has a smaller effect on the electronic structure of uranium than does chloride, since the \ce{U-Br} interaction has less of an effect on the  \ce{U-O_{yl}} bonding. Consistently, the $\nu_\mathrm{s}$ frequency is larger than that in \ce{[UO2Cl4]^{2-}} and is close to the bare uranyl value. For \ce{[UO2Br4]^{2-}} the shifts of vibrational frequencies between the ground and excited state are given in Table~S10 in the Supporting Information. 
 
\paragraph{Organic Cation Uranyl Tetrachloride: \ce{[{A336}]2[UO2Cl4]}.}

The \ce{[UO2Cl4]^{2-}} dianion in interaction with the organic ligand has been prepared under the conditions described in the Experimental Section. 
The final compound is predicted to be the uranyl tetrachloride associated with two extractant cations of methyltrioctylammonium \ce{[{A336}]2[UO2Cl4]} as a result of an ion-exchange mechanism~\cite{reillerA336termodynamics}. Interestingly, G\"{o}rller-Warland~{\etal}~\cite{B317002K} stabilized the \ce{[UO2Cl4]^{2-}} dianion in solution with tetrabutylammonium chloride (\ce{Bu4NCl}), a ligand belonging to the same group of tetraalkylammonium salts as Aliquate 336. For our purpose, the geometries of the ground and luminescent states of both complexes were optimized by placing the two countercations in the second coordination sphere in a $trans$ position. The bond distances are shown in Table~\ref{table:structure_complexes}, while the vibrational frequencies are shown in Table~\ref{table:freq}. 

If one compares the computed geometrical parameters to those of the models without an explicit second-coordination sphere (Table~\ref{table:structures_tetrahalides}), the countercations induce a weakening of the uranyl bond and loss of the $D_{4h}$ symmetry. These changes are due to the presence of weak hydrogen bonds between the hydrogen atoms of the alkyl chains of the cation with the oxygen and chloride atoms of the uranyl tetrachloride unit~\cite{sornein2006uranyl} (see Figure~S3 and Table~S3 in the Supporting Information). The ground-state geometries of uranyl tetrachloride in \ce{[Bu4N]2[UO2Cl4]} and \ce{[{A336}]2[UO2Cl4]} have been found to be nearly the same; the  \ce{U-O_{yl}} bond lengths are longer by about \SI{0.010}{\angstrom} than in the bare \ce{[UO2Cl4]^{2-}} complex. Moreover, because of the symmetry distortion and interaction of chloride ligands with the counterion, the \ce{U-Cl} distances differ by 0.026 up to \SI{0.118}{\angstrom} in comparison to the bare \ce{[UO2Cl4]^{2-}}. Even though our theoretical model does not account for the long-range effects induced by the presence of other species beyond the second coordination sphere, these gas-phase structures are in good accordance with the experimental crystal structures measured for parent compounds with a shorter alkyl chain, such as \ce{[Me4N]2[UO2Cl4]} and \ce{[Et4N]2[UO2Cl4]}~\cite{schnaars2013structural}. This observation makes us conclude that the chain length of a tetraalkylammonium cation has no influence on the \ce{U-O_{yl}} and \ce{U-Cl} bond-length values and that both organic countercations (\ce{R4N}) interact with the uranyl tetrachloride dianion in a similar way. Since the influence of long-range solvent effects on the bare \ce{[UO2Cl4]^{2-}} complexes is negligible, we rely in the following on the gas-phase structures for the \ce{[R4N]2[UO2Cl4]} complexes.

The addition of two countercations in our chemical model does not change the nature of the first excited state. As a result, the luminescent-state geometry is not significantly affected  by the countercations: in comparison to the bare \ce{[UO2Cl4]^{2-}} complexes, the  \ce{U-O_{yl}} bond length stretches by \SI{0.028}{\angstrom} for \ce{[R4N]2[UO2Cl4]} and the \ce{U-Cl} distances are from \num{0.003} to \SI{0.027}{\angstrom} longer for both types of complexes. 

\begin{table*}[!htbp]
	\centering
	\caption{Ground- and Excited-State Geometries of the \ce{[R4N]2[UO2Cl4]} Compounds Compared to Selected Experimental Results}
	\label{table:structure_complexes}
	\begin{tabular}{llllll}
		\toprule
		&$\mathrm{R_{U-O_{yl}}}$, \r{A}             & $\mathrm{R_{U-Cl}}$, \r{A}            & Compound                       & Method     & Ref.                                           \\ \hline
		Ground state&&&&&                                                                                                                                 \\ \hline
		& 1.766(6)        & 2.648(1)-2.677(1) & \ce{[Me4N]2[UO2Cl4]} & X-ray      & [\citenum{schnaars2013structural}] \\
		& 1.76(2)-1.77(3) & 2.65(1)-2.68(1)   & \ce{[Et4N]2[UO2Cl4]} & X-ray      & [\citenum{schnaars2013structural}] \\
		& 1.769           & 2.596-2.753       & \ce{[Bu4N]2[UO2Cl4]} & R-ECP/PBE0 & this work                                      \\
		& 1.767           & 2.609-2.740       & \ce{[{A336}]2[UO2Cl4]}  & R-ECP/PBE0 & this work                                      \\
		 \hline
		\multicolumn{6}{l}{Excited state}                                                                                                                                 \\ \hline
		& 1.796           & 2.599-2.780   & \ce{[Bu4N]2[UO2Cl4]} & R-ECP/PBE0 & this work                                      \\
		& 1.795           &      2.614-2.761    & \ce{[{A336}]2[UO2Cl4]}  & R-ECP/PBE0 & this work                                  \\
 \bottomrule   
	\end{tabular}
\end{table*}

The calculated frequencies have been compared with data from experimental Raman and IR measurements of different crystals or liquid samples of \ce{[R4N]2[UO2Cl4]}~\cite{schnaars2013structural, newbery1969vibrational,gal1984mid}. The computed symmetrical stretching frequency $\nu_\mathrm{s}$ for \ce{[Bu4N]2[UO2Cl4]} matches the value measured for the \ce{[Et4N]2[UO2Cl4]} crystal~\cite{newbery1969vibrational}. For \ce{[Bu4N]2[UO2Cl4]}, the gas-phase computed frequencies $\nu_\mathrm{U-Cl}$, $\nu_\mathrm{b}$, $\nu_\mathrm{s}$ and $\nu_\mathrm{a}$ are overestimated by 12, 24, 36 and \SI{26}{\per\cm}, respectively, with respect to the values measured in dichloromethane solution~\cite{gal1984mid}. From emission spectroscopy measurements the symmetrical stretching vibrational frequencies ($\nu_\mathrm{s}$) for \ce{[Bu4N]2[UO2Cl4]} in acetone and for \ce{[{A336}]2[UO2Cl4]} in $n$-dodecane were estimated as \num{823}~\cite{B317002K} and \SI{836}{\per\cm}, respectively. Our gas-phase calculations for these exact complexes yield slightly blue-shifted values: \num{869} and \SI{876}{\per\cm}. 

The differences in computed harmonic frequencies of \ce{[UO2Cl4]^{2-}} in the two \ce{[R4N]2[UO2Cl4]} complexes are found to be small. The counterions in the second coordination sphere only slightly affect the calculated frequencies with respect to the bare \ce{[UO2Cl4]^{2-}} ion; the $\nu_\mathrm{U-Cl}$ value is \SI{29}{\per\cm} higher and the $\nu_\mathrm{b}$ va;ie has increased by \SI{13}{\per\cm}, while the $\nu_\mathrm{s}$ and $\nu_\mathrm{a}$ modes were found to be lower by \num{18} and \SI{14}{\per\cm}, respectively. This is a result of the loss of symmetry together with the interplay of some motions of hydrogen and carbon atoms in alkyl chain in the vibrational motions. On the basis of the Franck-Condon principle, one can note that it should improve the band spacing between the vibronic progressions of theoretical luminescence spectrum that will be discussed below.

The shifts of theoretical frequencies between the ground and excited states of \ce{[{A336}]2[UO2Cl4]} are found to be similar to what was experimentally obtained for \ce{Cs2UO2Cl4}. The computed uranyl symmetrical stretching mode shifted by \SI{74}{\per\cm}, very close to the experimental value of \SI{82}{\per\cm}. However, the $\nu_\mathrm{U-Cl}$ and $\nu_\mathrm{b}$ values do not vary during the excitation. 

\subsection{Theoretical Absorption and Emission Energies}
As the luminescence spectrum of uranium (VI) complexes arises from the electronic transition from the lowest excited state to the ground state, coupled with the progression of vibronic bands, the examination of the whole electronic spectrum is pointless. In this step, we aim at foreseeing the sensitivity to the quantum chemical method of the first triplet excited state absorption and emission energies of the uranyl tetrachloride complex and of the spectral features of uranium~(VI) complexes. 

\begin{figure}[!htbp]
	\includegraphics[width=\linewidth]{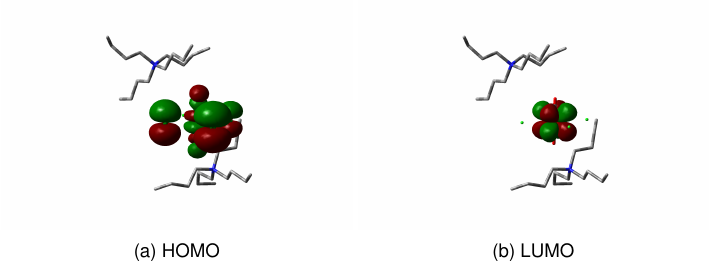}
	\caption{Highest occupied (HOMO) and lowest unoccupied molecular orbitals (LUMO) of \ce{[Bu4N]2[UO2Cl4]} in the gas phase obtained at the RECP DFT/PBE0 level of theory.}
	\label{fig:orbitals}
\end{figure}

The \ce{[UO2Cl4]^{2-}} dianion electronic spectrum has been computed previously with different levels of theory, and detailed discussions of the electronic structure can be found in the literature~\cite{doi:10.1063/1.2735297, doi:10.1063/1.2121608, doi:10.1021/jp003032h, doi:10.1021/jp3011266}. From previous studies it is known that for a uranyl dication coordinated by ligands the lowest excited state arises from the $\sigma_u$ highest molecular orbital (HOMO) to the $\delta_u$ lowest unoccupied orbital (LUMO). The HOMO corresponds to the bonding combination of uranium 5f and 6p oxygen 2p atomic orbitals and the 3p orbital of first-shell ligands, while the LUMO is a nonbonding uranium 5f orbital (Figure~\ref{fig:orbitals}). The detailed analysis of the atomic orbitals contributions performed with multireference CASSCF (complete active space self-consistent field) calculations by Pierloot and van Besien~\cite{doi:10.1063/1.2121608} suggests that the lowest-lying excitation corresponds to a metal-centered transition from the bonding to nonbonding orbital of uranium, with a marginal ligand-to-metal charge transfer character.  Hence, we can use in our discussion the \ce{UO2^{2+}} spin-free notation ($D_{\infty h}$), thus labeling the ground state as $^1\Sigma_g^+$ and the luminescent state as $^3\Delta_g$.

The vertical absorption ($\mathrm{E_{VA}}$) and emission ($\mathrm{E_{VE}}$) energies were obtained at the all-electron SOC CAM-B3LYP level of theory and are reported here together with experimental values (Table~\ref{tab:asb-emis}). They were computed with the summation of the fully relativistic electronic vertical energies associated with the spin-free zero-point energy correction. It should be noted that the experimental data correspond to the band-origin values obtained from polarized absorption or luminescence spectra and emission energies taken from UV-visible spectroscopic measurements. Thus, a direct comparison with theoretical results is not relevant, but rather we discuss whether the theoretical data reproduce the experimental trends.

\begin{table*}[!htbp]
	\centering
	\caption{Experimental and Computed Vertical Absorption ($\mathrm{E_{VA}}$) and Vertical Emission ($\mathrm{E_{VE}}$) Energies of Uranyl Tetrahalide Complexes (in \SI{}{\per\cm}). The computed values are obtained at the all-electron SOC CAM-B3LYP level of theory, and corrected with the spin-free Zero-Point Energy correction of the ground and luminescent states}
	\label{tab:asb-emis}

		\begin{tabular}{@{}lcccc@{}}
			\toprule
			\multicolumn{1}{c}{}            & \multicolumn{2}{c}{Theor.}                                                  & \multicolumn{2}{c}{Exp.} \\ \midrule
			\multicolumn{1}{c}{}            & $\mathrm{E_{VA}}$ & $\mathrm{E_{VE}}$                                                      & Band-origin  & Emission     \\ \hline
			\ce{[UO2Cl4]^{2-}} gas-phase & 20737        &19924&20096~$^a$& \\
			\ce{[UO2Cl4]^{2-}} acetone    & 20822          & 20116       &   &          \\
			\ce{[UO2Br4]^{2-}} gas-phase&    20746          & 20041& 19968~$^b$&                  \\
			\ce{[Bu4N]2[UO2Cl4]} acetone  &       & 20009&20097~$^c$ & 21000~$^c$  \\
			 \ce{[{A336}]2[UO2Cl4]} $n$-dodecane &      & 20041 &    & 21025    \\ 
			 \hline
			 \multicolumn{5}{l}{$^a$ Ref.~[\citenum{denning1976electronic}], $^b$ Ref.~[\citenum{flint1982electronic}], $^c$ Ref.~[\citenum{B317002K}]  } \\ \bottomrule		 
		\end{tabular}
\end{table*}
 
As seen in Table \ref{tab:asb-emis}, the vertical absorption energy of \ce{[UO2Cl4]^{2-}} increases by \SI{85}{\per\cm} with the addition of acetone due to its small polarity. A blue shift of \SI{117}{\per\cm} is observed when chlorides are replaced by bromide ligands. In contrast, the band origins seem red-shifted by \SI{128}{\per\cm} from chloride to bromide complexes according to experimental data~\cite{denning1976electronic, flint1982electronic}. The presence of multiple nonequivalent uranyl sites in the \ce{[UO2Br4]^{2-}} crystal studied by Flint~{\etal}~\cite{flint1982electronic} may cause a reverse shift of the whole spectrum and hamper a direct comparison between theoretical and experimental results. 
 
The luminescent state is known to be more sensitive to the solvent polarity~\cite{lakowiczbook}. We indeed observed a larger shift of about \SI{192}{\per\cm}  with the inclusion of acetone solvent in comparison to that for absorption energies. With addition of \ce{Bu4N} in the second coordination sphere, the $\mathrm{E_{VE}}$ increased by \SI{85}{\per\cm}, as expected since the displacements between the ground- and excited-state \ce{[UO2Cl4]^{2-}} geometries are larger with than without counter cations. Both theoretical $\mathrm{E_{VE}}$ and experimental emission energies of \ce{[UO2Cl4]^{2-}} are blue shifted by \num{32} or \SI{25}{\per\cm} on immersion in $n$-dodecane or acetone, respectively.

It is worth noting that in the experimental spectra, the band-origin values of \ce{[UO2Cl4]^{2-}} are the same within \SI{1}{\per\cm} in the \ce{Cs2UO2Cl4} crystal and in \ce{[Bu4N]2[UO2Cl4]} in acetone. This is fully consistent with the fact that the singly occupied molecular orbitals in the luminescent state do not show any contribution from the second-sphere counterions (See Figure~\ref{fig:orbitals}).

\subsection{Theoretical Luminescence Spectra}

To simplify the comparison within all spectra, those obtained from our quantum chemical calculations or from experiment, they are all normalized so that the first peak in the low-energy region matches the experimental amplitude. Neither homogeneous nor heterogeneous line broadening effects are included in our computed spectra because of their complicated prediction~\cite{doi:10.1021/ct200419x}. We will thus restrict ourselves to discussing the stick spectra.

\begin{figure}[!htbp]
	\centering
	\includegraphics[width=0.9\linewidth]{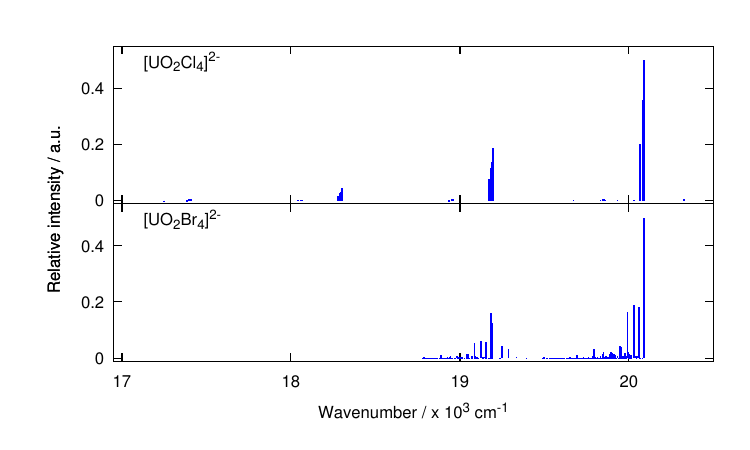}
	\caption{Theoretical luminescence spectra of \ce{[UO2Cl4]^{2-}} and \ce{[UO2Br4]^{2-}} in the gas phase computed at \SI{300}{\kelvin}.}
	\label{fig:first coord sphere}
\end{figure}

The most relevant parameters influencing the spectral intensity distribution are the vibrational wave functions of both initial and excited states. The spectral shape is significantly linked to the bond-length changes between the two electronic states, as identified by Su~{\etal}~\cite{su2011uranyl} thanks to a semiclassical vibronic approach. In that study, they solely considered the progression of one vibrational mode $\nu_\mathrm{s}$. They concluded that the computed bond-length of the excited state is estimated without a sufficient accuracy to properly simulate the intensities. In our study, the theoretical luminescence spectra of \ce{[UO2Cl4]^{2-}} and \ce{[UO2Br4]^{2-}} in gas phase were obtained by including all vibrational modes to be able to assign all the fine details of the measured vibrational progressions. The theoretical spectra are displayed in Figure~\ref{fig:first coord sphere}. The first band appearing in the progression corresponds to the 0-0 vibrational transition. In all uranium(VI) compounds at room temperature (\SI{300}{\kelvin}), luminescence arises from a low oscillator strength electronic transition followed by the vibrational progression in the $\nu_\mathrm{s}$ mode of the electronic ground state equal to \SI{894}{\per\cm}. Because $\nu_\mathrm{s}$ is totally symmetrical, it preserves the symmetry of the vibronic (electronic+vibrational) wave function. In \ce{[UO_2Cl_4]^{2-}}, the totally symmetrical $\nu_\mathrm{U-Cl}$ mode is also excited vibronically, therefore contributing to a line in the vibronic progression which is distant from the 0-0 line by \SI{235}{\per\cm}. However, as we already commented upon the fact that the \ce{U-X} lengthening from the ground and excited states is underestimated by our quantum approach, we cannot expect the computed relative intensities to match the experimental intensities. Still, our predicted spectra place the peak distribution and band spacing in great agreement with the experimental results. As other vibronic transitions might not be easily visible in Figure~\ref{fig:first coord sphere}, we provide a detailed assignment in the Supporting Information.

Setting the \ce{[UO2Cl4]^{2-}} spectra computed in the gas phase side-by-side with those computed in $n$-dodecane and acetone solvents (see Figue S2 in the Supporting Information) reveals that long-range solvent interactions have a minor influence on the spectral profiles, mostly on the intensities of $\nu_\mathrm{U-X}$ vibrational mode contributions. Nevertheless, the \ce{[UO2Br4]^{2-}} luminescence spectrum displays some more intense contributions and longer vibronic progressions in comparison to \ce{[UO2Cl4]^{2-}} (see the bottom panel on Figure~\ref{fig:first coord sphere}). This is due to a larger geometrical displacement between the ground- and excited- state structures (see Table~S10 in the Supporting Information), which leads to the appearance of \ce{U-Br} in-plane bending mode progression, distant by \SI{96}{\per\cm} from the 0-0 line.

\begin{figure}[!htbp]
	\centering
	\includegraphics[width=0.9\linewidth]{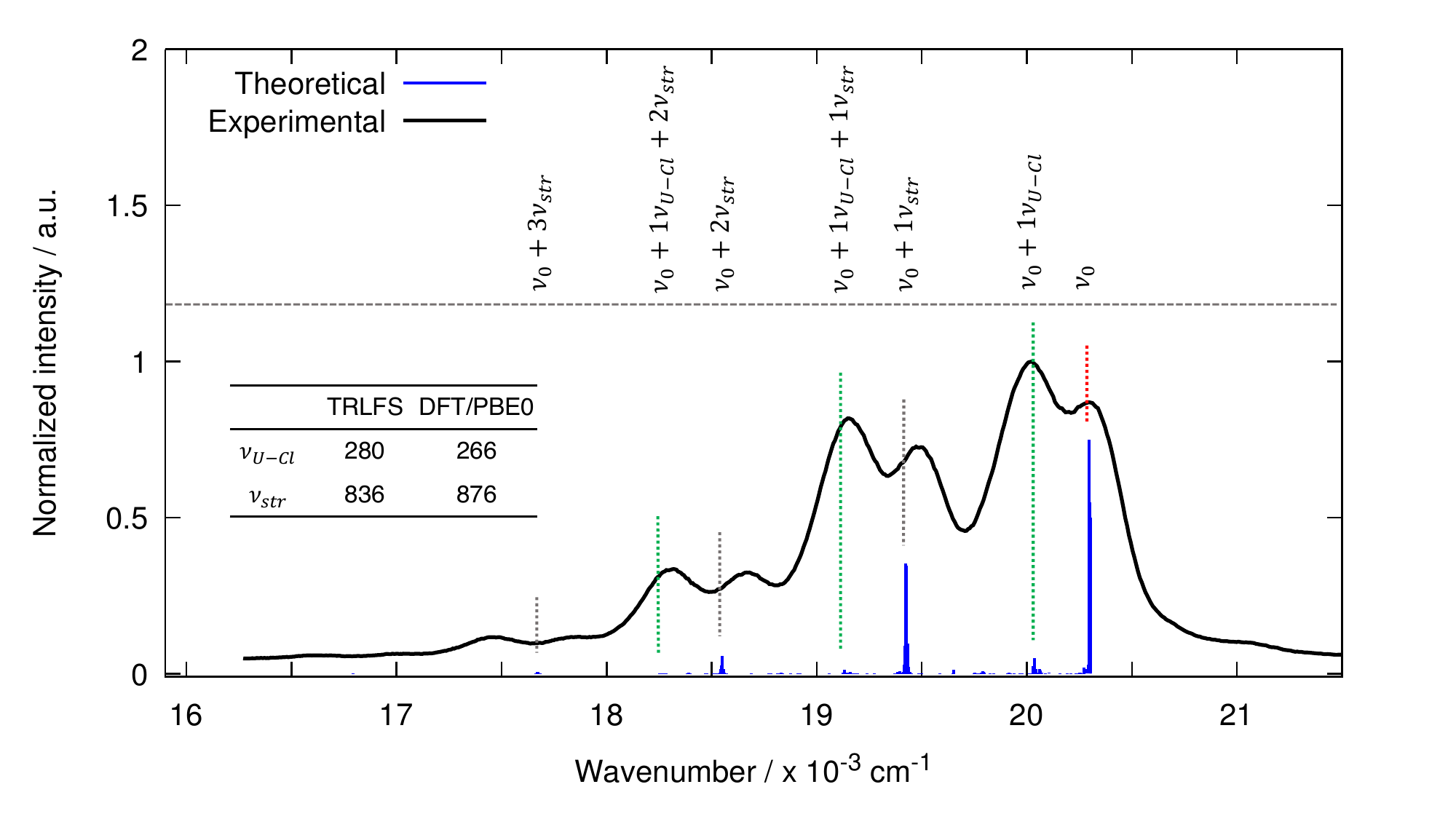}
	\caption{Experimental and theoretical luminescence spectra of \ce{[{A336}]2[UO2Cl4]} in $n$-dodecane and the gas phase, respectively at \SI{300}{\kelvin}.}
	\label{fig:a336-theory-exp}
\end{figure}

The computed luminescence spectrum of \ce{[{A336}]2[UO2Cl4]} in the gas phase is overlaid with the experimental spectrum in Figure~\ref{fig:a336-theory-exp}. This figure highlights the fact that both the energy positions and the band spacings of the computed vibronic transitions match nicely the experimental envelope, though with the reservation that the band spacing is somewhat red-shifted, as the ground-state symmetrical stretching is slightly overestimated by the R-ECP DFT/PBE0 level of theory and anharmonicity corrections are not accounted for. The vibronic progressions do not change significantly in comparison to bare \ce{[UO2Cl4]^{2-}}. In agreement with the results of Görller-Walrand {\etal}~\cite{B317002K}, our theoretical approach embeds four vibronic progressions (see Table~S8 in the Supporting Information), the first progression being from $\nu_\mathrm{s}$, which is in our calculation just \SI{40}{\per\cm} shifted in comparison to experiment, and the second from the rocking vibrational modes. While Görller-Walrand {\etal} attributed the other progressions to the out-of-plane bending of the chloride anions, our analysis assigns them to the symmetrical and antisymmetrical \ce{Cl-U-Cl} stretchings.
 
\section{Conclusions}

Probing and quantifying the long-range effects of large organic countercations on \ce{[UO2Cl4]^{2-}} complex luminescence spectra was realized by two approaches: one experimental using TRLFS and the second using \emph{ab initio}-based approaches. 
For the latter, relativistic DFT quantum chemical methods were found to be quantitative and effective tools to rationalize and predict uranium(VI)-based complex luminescence properties. Thanks to a benchmark with respect to previous studies on \ce{[UO2Cl4]^{2-}} in the crystal and solvent phases, we have been able to assess the degree of reliability of such an approach by a comparison of ground- and luminescent-states structures and frequencies. However, this theoretical protocol reaches some limitations, since we were not able to compute the exact geometrical displacements of the complexes upon excitation from the ground to the first excited state. As a result, the computed intensities of the vibronic bands do not coincide with the experimental intensities. Conversely, the peak positions of the luminescence spectra are correctly reproduced, and the band spacings and theoretical assignments are in good accordance with our experimental data. 

In this work, we have used stepwise growth chemical models to quantify the influence of (i) the first coordination sphere by substituting chloride anions by bromides, (ii) the influence of second-sphere countercations, and (iii) long-range solvent effects with a polarizable continuum model, with the aim of approaching the experimental conditions. We have found that both long-range solvent effects and the second coordination sphere have little influence on the vibronic intensities, while their effect is more significant for the prediction of other spectroscopic parameters, such as emission energies and vibronic band spacings of the \ce{[UO2Cl4]^{2-}} complex. This also confirms that TRLFS spectroscopy may not be sensitive enough to discriminate long-range interactions induced by the countercations present in the vicinity of the luminescent center. 

\section*{Acknowledgements}
We acknowledge support by the French government through the Program "Investissement d'avenir" (LABEX CaPPA / ANR-11-LABX-0005-01 and I-SITE ULNE / ANR-16-IDEX-0004 ULNE), as well as by the Ministry of Higher Education and Research, Hauts de France council and European Regional Development Fund (ERDF) through the Contrat de Projets État-Région (CPER CLIMIBIO). Furthermore, this work was granted access to the HPC resources of [CINES/IDRIS/TGCC] under the allocation 2018-2019 [A0050801859] made by GENCI. We also acknowledge the CEA for the Ph.D. grant given to H.O.

\section*{Associated content}
The following file supplemental-uo2x4-Oher.pdf is available free of charge. It contains:
\begin{itemize}
  \item Theoretical spectra of the \ce{[UO2Cl4]^{2-}} obtained with the long-range counterion and solvent effects.
  \item Assignments for theoretical and experimental spectra.
  \item Cartesian coordinates for the calculated structures.
\end{itemize}

\newpage
\bibliography{uo2x4.bib}

\end{document}


\tableofcontents
\clearpage
\listoffigures

\clearpage
\listoftables
\clearpage

\section{Selection of the delay time}

\begin{figure}[!htbp]
	\centering
	\includegraphics[width=.8\textwidth]{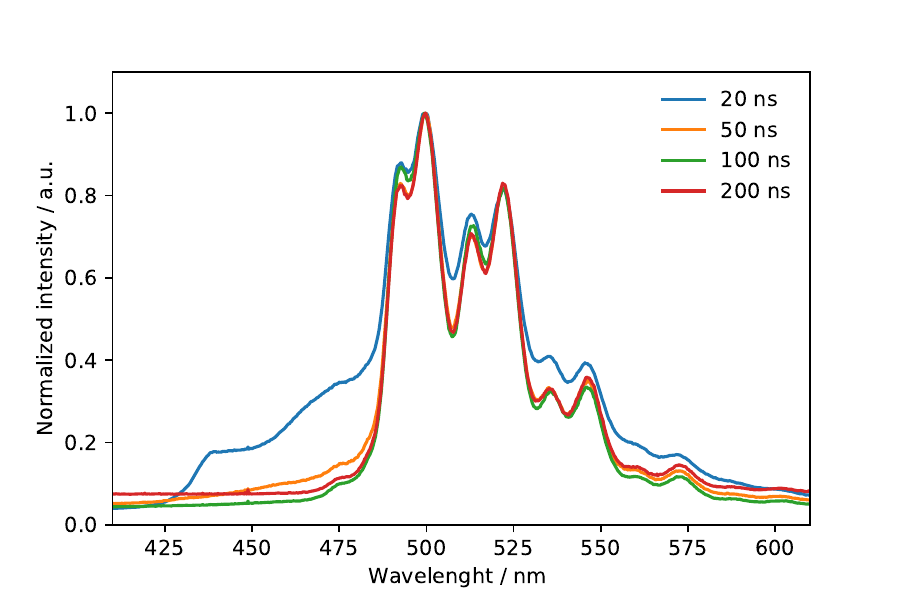}
	\caption{Normalized time-resolved luminescence spectra of \ce{[{A336}]_2[UO_2Cl_4]} in  $n$-dodecane measured at different delay time.}
	\label{fig:delay-time}
\end{figure}

To select a correct delay time for recording of time-resolved spectrum of \ce{[{A336}]_2[UO_2Cl_4]} in  $n$-dodecane, we have performed measurements at several delay time, 20, 50, 100 and \SI{200}{\ns}. The corresponding recorded spectra are shown in Figure \ref{fig:delay-time}.

\FloatBarrier
\FloatBarrier

\begin{figure}[!h]
	\begin{minipage}[t]{0.48\linewidth}	
		\includegraphics[width=\linewidth]{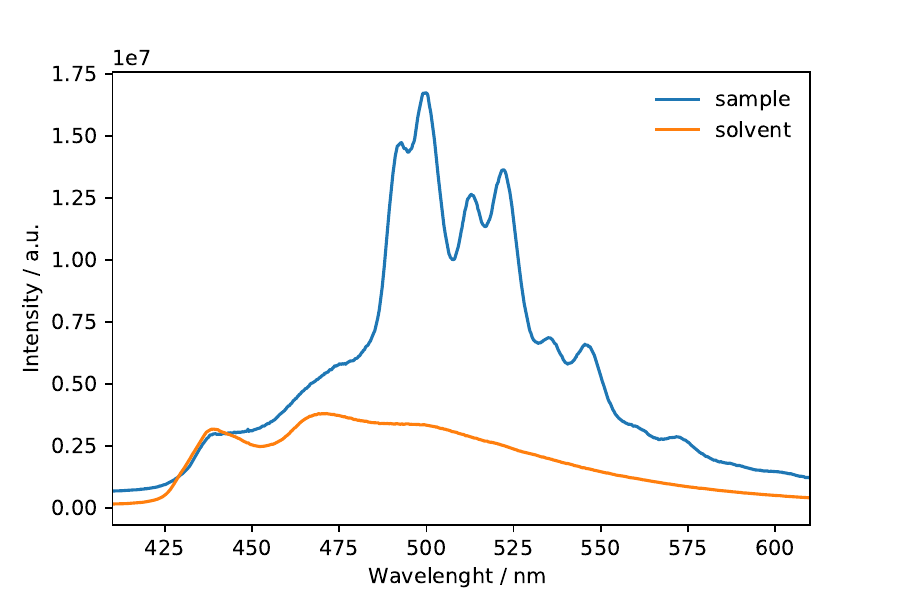}
		\subcaption{\SI{20}{ns}}
	\end{minipage}
	\begin{minipage}[t]{0.48\linewidth}
		\includegraphics[width=\linewidth]{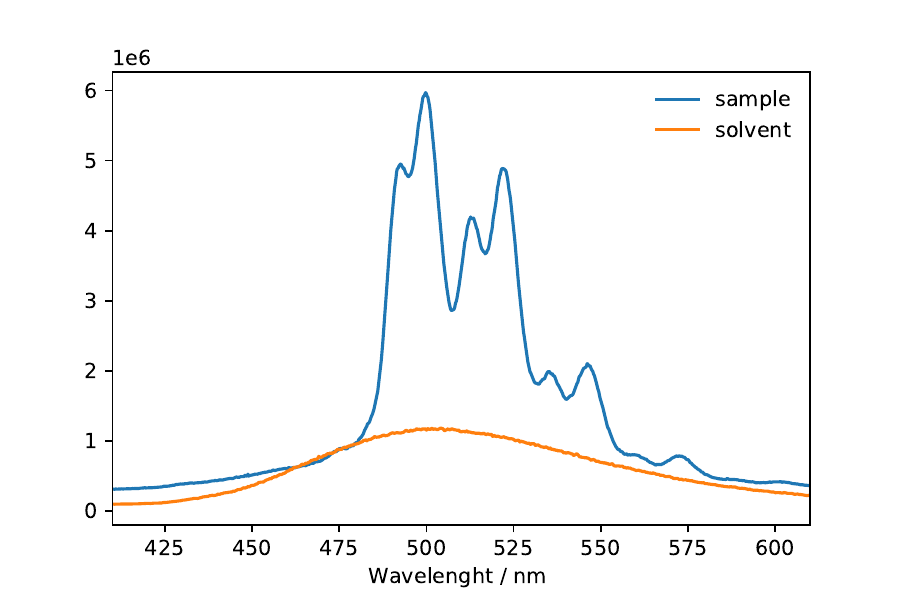}
		\subcaption{\SI{50}{ns}}
	\end{minipage}
	\caption{Comparison of reference sample(without U) with the sample of \ce{[{A336}]_2[UO_2Cl_4]} in  $n$-dodecane measured at different delay time.}
	\label{fig:20-50ns}
\end{figure}

By inspecting the normalized spectra at different delay times, we have found that positions of maxima remain the same, along with the vibronic features. However, at short delay times (20 and \SI{50}{\ns}), a background luminescence contribution appears in the short  wavelength range (around \SI{430}{\nm}), which, by comparison to the spectra of the reference media (solution without uranium), might be attributed to the luminescence of the solvent or the A336 molecule (Figure~\ref{fig:20-50ns}), with its own exponential decay. 

This background emission vanished for larger delay times (100, up to \SI{200}{\ns}), and as the recorded spectra are essentially the same for these two-time values, we chose the \SI{100}{\ns} delay time. 

\clearpage
\section{Supplemental Tables}
\FloatBarrier
	\begin{table}[!htbp]
		\centering
		\caption{Experimental luminescence data of the \ce{[R_4N]_2[UO_2Cl_4]} complexes at room temperature. The  $\nu$ corresponds to the position of the band, the $\Delta \nu$ is a difference between the bands position in a region, the $\Delta \nu_n $ is a difference between the peaks of the same nature located in different regions. All data are in \si{\per\cm}. The luminescence spectra are shown on Figure 2 in the main text.}
		\label{tab:my-table}
		\begin{tabular}{cc|cccc|cccc}
			\toprule
			&&\multicolumn{4}{c}{\ce{[{A336}]_2[UO_2Cl_4]} in  $n$-dodecane}      & \multicolumn{4}{|c}{\ce{[Bu_4N]_2[UO_2Cl_4]} in acetone} \\ \hline
			&     &   $\nu / cm^{-1}$    & $\Delta \nu$ & $\Delta \nu_n $ & Intensity &  $\nu / cm^{-1}$    & $\Delta \nu$ & $\Delta \nu_n $   & Intensity   \\ \hline
			Region &     & 21025 &            &               & 0.06    &         &              &                 &             \\ \hline
			1      & $I_1$  & 20295 & 730        &               & 0.84      & 20325   & -            &                 & 0.74        \\
			& $I'_1$ & 20015 & 280        &               & 1.00      & 20032   & 293          &                 & 0.95        \\
			& \multirow{2}{*}{$I^{*}_1$} &       &            &               &           & 19944   & 88           &                 & 0.91        \\
			&     &       &            &               &           & 19763   & 181          &                 & 0.40        \\ \hline
			2      & $I_2$  & 19488 & 527        & 807           & 0.70      & 19501   & 262          & 824             & 0.84        \\
			& $I'_2$ & 19155 & 333        & 860           & 0.81      & 19238   & 263          & 794             & 0.83        \\
			& \multirow{2}{*}{$I^*_2$} &       &            &               &           & 19113   & 125          & 831             & 1.00        \\
			&     &       &            &               &           & 18925   & 188          & 838             & 0.34        \\ \hline
			3      & $I_3$  & 18651 & 504        & 837           & 0.28      & 18664   & 261          & 837             & 0.39        \\
			& $I'_3$ & 18305 & 346        & 850           & 0.30      & 18403   & 261          & 836             & 0.36        \\
			& \multirow{2}{*}{$I^*_3$} &       &            &               &           & 18282   & 121          & 832             & 0.51        \\
			&     &       &            &               &           & 18103   & 179          & 822             & 0.16        \\ \hline
			4      & $I_4$  & 17839 & 466        & 812           & 0.08      & 17825   & 278          & 838             & 0.11        \\
			& $I'_4$ & 17454 & 385        & 851           & 0.08      & 17593   & 232          & 810             & 0.09        \\
			& \multirow{2}{*}{$I^*_4$}&       &            &               &           & 17452   & 141          & 830             & 0.16        \\
			&     &       &            &               &           & 17289   & 163          & 814             & 0.05        \\ \hline
			5      & $I_5$  & 16996 & 458        & 843           & 0.02      & 17013   & 277          & 813             & 0.03        \\
			& $I'_5$ & 16625 & 371        & 829           & 0.02      & 16795   & 217          & 798             & 0.02        \\
			& \multirow{2}{*}{$I^*_5$} &       &            &               &           & 16628   & 168          & 824             & 0.03        \\
			&     &       &            &               &           & 16442   & 186          & 847             & 0.01       \\ \bottomrule
		\end{tabular}
	\end{table}
	\FloatBarrier
	
\clearpage

	\FloatBarrier
	\begin{table}[!ht]
		\centering
		\caption{Ground and excited state harmonic and anharmonic frequencies (in \si{\per\cm}) of the \ce{[UO2Cl4]^{2-}} complex in a gas phase computed by G16 at R-ECP DFT/PBE0 and TD-DFT/PBE0 level of theory for the ground and first excited states respectively.}
		\label{tab:harm-anharm-freq}
		\begin{tabular}{ccc|ccl}
			\hline
			& \multicolumn{2}{c}{Ground state} & \multicolumn{2}{c}{Excited state} & \multicolumn{1}{c}{\multirow{2}{*}{Nature of the band}} \\ \cline{1-5}
			$\nu$& $E_{harm}$       & $E_{anharm}$    &  $E_{harm}$       & $E_{anharm}$ & \multicolumn{1}{c}{}                                 \\ \hline
			1  & 75.7     & 76.4     & 71.3     & 72.4    & out-of-plane asymmetrical Cl-U-Cl bending            \\ \hline
			2  & 76.6     & 76.6     & 77.2     & 77.2   & \multirow{2}{*}{in plane Cl-U-Cl bending}               \\ 
			3  & 76.6     & 76.6     & 77.2     & 77.2    &                                                      \\ \hline
			4  & 97.3     & 98.0    & 98.7     & 99.8    & U-Cl scissoring                                       \\ \hline
			5  & 110.0    & 110.3   & 104.1    & 105.0   & out-of-plane symmetrical Cl-U-Cl bending                \\ \hline
			6  & 197.2    & 193.7   & 170.5    & 166.7   & \multirow{2}{*}{\ce{O_{yl}-U-O_{yl}} rocking}                         \\ 
			7  & 197.2    & 193.7   & 170.5    & 166.7   &                                                      \\ \hline
			8  & 202.3    & 202.5   & 199.2    & 199.0   & asymmetrical Cl-U-Cl stretching                     \\ \hline
			9  & 219.1      & 219.2   & 214.5     & 214.6   & \multirow{2}{*}{asymmetrical Cl-U-Cl stretching}                           \\ 
			10 & 219.1      & 219.2   & 214.5     & 214.6   &                                                      \\ \hline
			11 & 237.3    & 237.0   & 232.6     & 231.6    & symmetrical Cl-U-Cl  stretching                                    \\ \hline
			12 & 274.5     & 271.4   & 271.8    & 268.5   & \multirow{2}{*}{\ce{O_{yl}-U-O_{yl}} bending}                         \\ 
			13 & 274.5     & 271.4   & 271.8    & 268.5   &                                                      \\ \hline
			14 & 894.5    & 889.4   & 813.7    & 810.9    & symmetrical \ce{O_{yl}-U-O_{yl}} stretching                                          \\ \hline
			15 & 974.2    & 966.9   & 858.9    & 854.3   & asymmetrical \ce{O_{yl}-U-O_{yl}} stretching                                      \\ \hline
		\end{tabular}
	\end{table}
	\FloatBarrier
	
	\clearpage
	
	\FloatBarrier
	\begin{table}[!htbp]
		\centering
		\caption{The chloride-hydrogen bond lengths (in \r{A}) in \ce{[R4N]2[UO2Cl4]} complexes computed at the R-ECP DFT/PBE0 and TD-DFT/PBE0 levels of theory for the ground and first excited states, respectively. Atom labeling corresponds to the one shown on Figure~\ref{fig:hyd-bond-bu4n}.}
		\label{tab:hyd-bond}
		\begin{tabular}{ccccccc}
			\toprule
			\multicolumn{3}{c}{\ce{[Bu4N]2[UO2Cl4]}} &  & \multicolumn{3}{c}{\ce{[{A336}]2[UO2Cl4]}} \\ \hline
			& GS           & ES          &  &                 & GS          & ES          \\ \hline
			Cl33-H83        & 2.534        & 2.524       &  & Cl25-H57        & 2.427       & 2.411       \\
			Cl36-H53        & 2.642        & 2.606       &  & Cl28-H48        & 2.557       & 2.535       \\
			Cl33-H97        & 2.646        & 2.632       &  & Cl28-H37        & 2.656       & 2.626       \\
			Cl36-H71        & 2.789        & 2.744       &  & Cl28-H151       & 2.672       & 2.685       \\
			Cl36-H66        & 2.893        & 2.885       &  & Cl27-H71        & 2.851       & 2.901      \\ \bottomrule
		\end{tabular}
	\end{table}
	\FloatBarrier

\clearpage

\FloatBarrier
\begin{table}[!htbp]
	\centering
	\caption{Assignment of the \ce{[UO2Cl4]^{2-}} (gas-phase) theoretical luminescence spectrum. The energy of the spectrum was adjusted to experimental band-origin value of \ce{[UO2Cl4]^{2-}} in acetone. The nature of bands is explained in Table \ref{tab:harm-anharm-freq}.}
	\label{tab:uo2cl42-gas-g16-ass}
	\begin{tabular}{l|ccc|l}
		\toprule
		Region & E, \si{\per\cm} & $\Delta \nu$, \si{\per\cm}& I, a.u. & Identification                  \\
		\hline
		0      & 20325.0 &     |     & 6.50E-01   & 0(0)-\textgreater{}1(0)         \\
		& 20087.9 & 237.1    & 1.43E-02   & 0(0)-\textgreater{}1($1\nu_{11}$)      \\
		& 19930.6 & 394.4    & 1.72E-03   & 0(0)-\textgreater{}1($2\nu_{7}$)       \\
		& 19850.8 & 474.2    & 3.29E-04   & 0(0)-\textgreater{}1($2\nu_{11}$)      \\ \hline
		1      & 19430.6 & 894.4    & 2.46E-01   & 0(0)-\textgreater{}1($1\nu_{14}$)      \\
		& 19193.5 & 237.1    & 7.21E-03   & 0(0)-\textgreater{}1($1\nu_{11}$, $1\nu_{14}$ ) \\
		& 19036.2 & 394.4    & 6.52E-04   & 0(0)-\textgreater{}1($2\nu_{7}$, $1\nu_{14}$)  \\
		& 18955.6 & 475.0    & 1.84E-04   & 0(0)-\textgreater{}1($2\nu_{11}$,$1\nu_{14}$) \\ \hline
		2      & 18536.2 & 894.4    & 5.91E-02   & 0(0)-\textgreater{}1($2\nu_{14}$)      \\
		& 18299.1 & 237.1    & 2.11E-03   & 0(0)-\textgreater{}1($1\nu_{11}$ , $2\nu_{14}$) \\
		& 18141.8 & 394.4    & 1.56E-04   & 0(0)-\textgreater{}1($2\nu_{7}$, $2\nu_{14}$)  \\
		& 18061.2 & 475.0    & 5.94E-05   & 0(0)-\textgreater{}1($2\nu_{11}$,$2\nu_{14}$) \\ \hline
		3      & 17641.8 & 894.4    & 1.11E-02   & 0(0)-\textgreater{}1($3\nu_{14}$)      \\
		& 17404.7 & 237.1    & 4.66E-04   & 0(0)-\textgreater{}1($1\nu_{11}$, $3\nu_{14}$) \\
		& 17247.4 & 394.4    & 2.95E-05   & 0(0)-\textgreater{}1($2\nu_{7}$,$3\nu_{14}$)  \\
		& 17166.7 & 475.0    & 1.44E-05   & 0(0)-\textgreater{}1($2\nu_{11}$, $3\nu_{14}$) \\ \hline
		4      & 16747.4 & 894.4    & 1.80E-03   & 0(0)-\textgreater{}1($4\nu_{14}$)      \\ 
		& 16510.3 & 237.1    & 8.61E-05   & 0(0)-\textgreater{}1($1\nu_{11}$,$4\nu_{14}$) \\\hline
		5      & 15853.0 & 894.4    & 2.62E-04   & 0(0)-\textgreater{}1($5\nu_{14}$)     \\ \bottomrule
	\end{tabular}
\end{table}
\FloatBarrier

\clearpage

\FloatBarrier
\begin{table}[!htbp]
	\centering
	\caption{Assignment of the \ce{[UO2Cl4]^{2-}} (dodecane) theoretical luminescence spectrum. The energy of the spectrum was adjusted to experimental band-origin value of \ce{[UO2Cl4]^{2-}} in acetone. The nature of bands is explained in Table \ref{tab:harm-anharm-freq}.}
	\label{tab:uo2cl42-dod-ass}
	\begin{tabular}{l|ccc|l}
		\toprule
		Region & E, \si{\per\cm} & $\Delta \nu$, \si{\per\cm}& I, a.u. & Identification                  \\
		\hline
		0      & 20325.0 &       & 6.59E-01 & 0(0)-\textgreater{}1(0)         \\
		& 20116.9 & 208.1 & 1.48E-04 & 0(0)-\textgreater{}1($2\nu_{5}$)       \\
		& 20085.5 & 239.5 & 2.42E-02 & 0(0)-\textgreater{}1($1\nu_{11}$)      \\
		& 19926.6 & 398.4 & 1.79E-03 & 0(0)-\textgreater{}1($2\nu_{7}$)       \\
		& 19845.9 & 479.1 & 6.63E-04 & 0(0)-\textgreater{}1($2\nu_{11}$)      \\ \hline
		1      & 19435.4 & 889.6 & 2.34E-01 & 0(0)-\textgreater{}1($1\nu_{14}$)      \\
		& 19227.4 & 208.1 & 5.26E-05 & 0(0)-\textgreater{}1($2\nu_{5}$, $1\nu_{14}$)  \\
		& 19195.9 & 239.5 & 8.19E-03 & 0(0)-\textgreater{}1($1\nu_{11}$, $1\nu_{14}$) \\
		& 19037.0 & 398.4 & 6.34E-04 & 0(0)-\textgreater{}1($2\nu_{7}$, $1\nu_{14}$)  \\
		& 18956.4 & 479.1 & 2.17E-04 & 0(0)-\textgreater{}1($2\nu_{11}$, $1\nu_{14}$) \\ \hline
		2      & 18545.9 & 889.6 & 5.30E-02 & 0(0)-\textgreater{}1($2\nu_{14}$)      \\
		& 18337.8 & 208.1 & 1.19E-05 & 0(0)-\textgreater{}1($2\nu_{5}$, $2\nu_{14}$)  \\
		& 18306.3 & 239.5 & 1.79E-03 & 0(0)-\textgreater{}1($1\nu_{11}$, $2\nu_{14}$) \\
		& 18147.5 & 398.4 & 1.44E-04 & 0(0)-\textgreater{}1($2\nu_{7}$, $2\nu_{14}$)  \\
		& 18066.8 & 479.1 & 4.62E-05 & 0(0)-\textgreater{}1($2\nu_{11}$, $2\nu_{14}$) \\ \hline
		3      & 17656.3 & 889.6 & 9.47E-03 & 0(0)-\textgreater{}1($3\nu_{14}$)      \\
		& 17416.8 & 239.5 & 3.09E-04 & 0(0)-\textgreater{}1($1\nu_{11}$, $3\nu_{14}$) \\
		& 17257.9 & 398.4 & 2.57E-05 & 0(0)-\textgreater{}1($2\nu_{7}$, $3\nu_{14}$)  \\ \hline
		4      & 16766.7 & 889.6 & 1.46E-03 & 0(0)-\textgreater{}1($4\nu_{14}$)      \\
		& 16527.2 & 239.5 & 4.60E-05 & 0(0)-\textgreater{}1($1\nu_{11}$, $4\nu_{14}$) \\ \hline
		5      & 15877.2 & 889.6 & 2.01E-04 & 0(0)-\textgreater{}1($5\nu_{14}$)     \\ \bottomrule
	\end{tabular}
\end{table}
\FloatBarrier

\clearpage

\FloatBarrier
\begin{table}[!htbp]
	\centering
	\caption{Assignment of the  \ce{[UO2Cl4]^{2-}} (acetone) theoretical luminescence spectrum. The energy of the spectrum was adjusted to experimental band-origin value of \ce{[UO2Cl4]^{2-}} in acetone. The nature of bands is explained in Table \ref{tab:harm-anharm-freq}.}
	\label{tab:uo2cl42-acetone-ass}
	\begin{tabular}{l|ccc|l}
		\toprule
		Region & E, \si{\per\cm} & $\Delta \nu$, \si{\per\cm}& I, a.u. & Identification                  \\
		\hline
		0      & 20325.0 &          & 5.30E-01   & 0(0)-\textgreater{}1(0)         \\
		& 20081.4 & 243.6    & 1.45E-01   & 0(0)-\textgreater{}1($1\nu_{11}$)      \\
		& 19929.0 & 396.0    & 1.56E-03   & 0(0)-\textgreater{}1($2\nu_{7}$)       \\
		& 19838.7 & 486.3    & 2.08E-02   & 0(0)-\textgreater{}1($2\nu_{11}$)      \\
		& 19595.1 & 729.9    & 2.10E-03   & 0(0)-\textgreater{}1($3\nu_{11}$)      \\ \hline
		1      & 19441.9 & 883.1    & 1.79E-01   & 0(0)-\textgreater{}1($1\nu_{14}$)      \\
		& 19198.3 & 243.6    & 4.82E-02   & 0(0)-\textgreater{}1($1\nu_{11}$, $1\nu_{14}$) \\
		& 19045.9 & 396.0    & 5.28E-04   & 0(0)-\textgreater{}1( $2\nu_{7}$, $1\nu_{14}$)  \\
		& 18955.6 & 486.3    & 6.82E-03   & 0(0)-\textgreater{}1($2\nu_{11}$, $1\nu_{14}$) \\ \hline
		2      & 18558.8 & 883.1    & 3.89E-02   & 0(0)-\textgreater{}1( $1\nu_{14}$)      \\
		& 18316.0 & 242.8    & 1.03E-02   & 0(0)-\textgreater{}1($1\nu_{11}$, $2\nu_{14}$) \\
		& 18162.8 & 396.0    & 1.14E-04   & 0(0)-\textgreater{}1($2\nu_{7}$, $2\nu_{14}$)  \\
		& 18072.4 & 486.3    & 1.44E-03   & 0(0)-\textgreater{}1($2\nu_{11}$, $2\nu_{14}$) \\ \hline
		3      & 17675.6 & 883.1    & 6.66E-03   & 0(0)-\textgreater{}1($3\nu_{14}$)      \\
		& 17432.9 & 242.8    & 1.74E-03   & 0(0)-\textgreater{}1($1\nu_{11}$, $3\nu_{14}$) \\
		& 17280.5 & 395.2    & 1.96E-05   & 0(0)-\textgreater{}1($2\nu_{7}$, $3\nu_{14}$)  \\
		& 17189.3 & 486.3    & 2.40E-04   & 0(0)-\textgreater{}1($2\nu_{11}$, $3\nu_{14}$) \\ \hline
		4      & 16793.3 & 882.3    & 9.80E-04   & 0(0)-\textgreater{}1($4\nu_{14}$)      \\
		& 16549.8 & 243.6    & 2.54E-04   & 0(0)-\textgreater{}1($1\nu_{11}$, $4\nu_{14}$) \\ \hline
		5      & 15910.2 & 883.1    & 1.30E-04   & 0(0)-\textgreater{}1($5\nu_{14}$) \\ \bottomrule
	\end{tabular}
\end{table}
\FloatBarrier

\clearpage

\FloatBarrier
\begin{table}[!htbp]
	\centering
	\caption{Assignment of the \ce{[UO2Br4]^{2-}} theoretical luminescence spectrum. The energy of the spectrum was adjusted to experimental band-origin value of \ce{[UO2Cl4]^{2-}} in acetone. The bands representation is taken from the \ce{[UO2Cl4]^{2-}} frequencies, the nature of bands is explained in Table \ref{tab:harm-anharm-freq}.}
	\label{tab:uo2br42-gas-ass}
	\begin{tabular}{l|ccc|l}
		\toprule
		Region & E, \si{\per\cm} & $\Delta \nu$, \si{\per\cm}& I, a.u. & Identification                  \\
		\hline
		0      & 20325.0 &          & 4.70E-01   & 0(0)-\textgreater{}1(0)         \\
		& 20229.0 & 96.0     & 1.63E-02   & 0(0)-\textgreater{}1($2\nu_{3}$)       \\
		& 20181.4 & 143.6    & 4.06E-02   & 0(0)-\textgreater{}1($1\nu_{11}$)      \\
		& 20081.4 & 243.6    & 7.03E-03   & 0(0)-\textgreater{}1($2\nu_{8}$)       \\
		& 19971.8 & 353.2    & 2.07E-03   & 0(0)-\textgreater{}1($2\nu_{7}$)      \\ \hline
		1      & 19418.5 & 906.5    & 1.53E-01   & 0(0)-\textgreater{}1($1\nu_{14}$)      \\
		& 19274.1 & 144.4    & 1.54E-02   & 0(0)-\textgreater{}1($1\nu_{11}$, $1\nu_{14}$) \\
		& 19174.9 & 243.6    & 2.29E-03   & 0(0)-\textgreater{}1($2\nu_{8}$,$1\nu_{14}$)  \\ \hline
		2      & 18511.2 & 907.3    & 2.76E-02   & 0(0)-\textgreater{}1($2\nu_{14}$)      \\
		& 18367.6 & 143.6    & 3.16E-03   & 0(0)-\textgreater{}1($1\nu_{11}$, $2\nu_{14}$) \\
		& 18268.4 & 242.8    & 4.13E-04   & 0(0)-\textgreater{}1($2\nu_{8}$,$2\nu_{14}$)  \\ \hline
		3      & 17604.7 & 906.5    & 3.63E-03   & 0(0)-\textgreater{}1($3\nu_{14}$)      \\
		& 17461.1 & 143.6    & 4.66E-04   & 0(0)-\textgreater{}1($1\nu_{11}$, $3\nu_{14}$) \\ \hline
		4      & 16697.4 & 907.3    & 3.88E-04   & 0(0)-\textgreater{}1($4\nu_{14}$)      \\
		& 16553.8 & 143.6    & 5.49E-05   & 0(0)-\textgreater{}1($1\nu_{11}$, $4\nu_{14}$) \\ \hline
		5      & 15790.9 & 906.5    & 3.56E-05   & 0(0)-\textgreater{}1($5\nu_{14}$)     \\ \bottomrule
	\end{tabular}
\end{table}
\FloatBarrier

\clearpage

\FloatBarrier
\begin{table}[!hpbp]
	\centering
	\caption{Assignment of the \ce{[Bu4N]2[UO2Cl4]} (gas-phase) theoretical luminescence spectrum. The energy of the spectrum was adjusted to experimental band-origin value of \ce{[UO2Cl4]^{2-}} in acetone. The nature of bands is shown in Table \ref{tab:harm-anharm-freq}.}
	\label{tab:uo2cl42-bu4n-ass}
	\begin{tabular}{l|ccc|l}
		\toprule
		Region & E, \si{\per\cm} & $\Delta \nu$, \si{\per\cm}& I, a.u. & Identification                  \\
		\hline
		0      & 20325   &          & 1.73E-01   & 0(0)-\textgreater{}1(0)         \\
		& 20114   & 211      & 1.24E-04   & 0(0)-\textgreater{}1($1\nu_{7}$)       \\
		& 20092   & 233      & 2.29E-03   & 0(0)-\textgreater{}1($1\nu_{8}$)       \\
		& 20056   & 269      & 1.32E-03   & 0(0)-\textgreater{}1($1\nu_{11}$)      \\
		& 19902   & 423      & 4.00E-04   & 0(0)-\textgreater{}1($2\nu_{7}$)       \\ \hline
		1      & 19456   & 869      & 5.26E-02   & 0(0)-\textgreater{}1($1\nu_{14}$)      \\
		& 19245   & 211      & 1.47E-05   & 0(0)-\textgreater{}1($1\nu_{7}$,$1\nu_{14}$)  \\
		& 19223   & 233      & 7.05E-04   & 0(0)-\textgreater{}1($1\nu_{8}$,$1\nu_{14}$)  \\
		& 19187   & 269      & 3.53E-04   & 0(0)-\textgreater{}1($1\nu_{11}$,$1\nu_{14}$) \\
		& 19034   & 423      & 1.19E-04   & 0(0)-\textgreater{}1($2\nu_{7}$,$1\nu_{14}$)  \\ \hline
		2      & 18588   & 869      & 1.03E-02   & 0(0)-\textgreater{}1($2\nu_{14}$)      \\
		& 18356   & 232      & 1.40E-04   & 0(0)-\textgreater{}1($1\nu_{8}$,$2\nu_{14}$)  \\
		& 18318   & 269      & 6.24E-05   & 0(0)-\textgreater{}1($1\nu_{11}$,$2\nu_{14}$) \\
		& 18165   & 423      & 2.32E-05   & 0(0)-\textgreater{}1($2\nu_{7}$,$2\nu_{14}$)  \\ \hline
		3      & 17719   & 869      & 1.60E-03   & 0(0)-\textgreater{}1($3\nu_{14}$)      \\
		& 17487   & 232      & 2.19E-05   & 0(0)-\textgreater{}1($1\nu_{8}$,$3\nu_{14}$)  \\ \hline
		4      & 16851   & 869      & 2.14E-04   & 0(0)-\textgreater{}1($4\nu_{14}$)      \\ \hline
		5      & 15982   & 869      & 2.57E-05   & 0(0)-\textgreater{}1($5\nu_{14}$)    \\
		\bottomrule 
	\end{tabular}
\end{table}
\FloatBarrier

\clearpage

\FloatBarrier
\begin{table}[!htbp]
	\centering
	\caption{Assignment of the \ce{[{A336}]2[UO2Cl4]} gas-phase theoretical luminescence spectrum. The energy of the spectrum was adjusted to experimental band-origin value of \ce{[{A336}]2[UO2Cl4]} in $n$-dodecane. The bands representation is taken from the \ce{[UO2Cl4]^{2-}} frequencies, with details found in Table \ref{tab:harm-anharm-freq}.}
	\label{tab:uo2cl42-a336-ass}
	\begin{tabular}{l|ccc|l}
		\toprule
		Region & E, \si{\per\cm} & $\Delta \nu$, \si{\per\cm}& I, a.u. & Identification                  \\
		\hline
		0      & 21025.0 &          & 7.55E-01 & 0(0)-\textgreater{}1(0)            \\
		& 20786.3 & 238.7    & 2.59E-03 & 0(0)-\textgreater{}1($1\nu_{8}$, $1\nu_{12}$)        \\
		& 20758.9 & 266.1    & 1.25E-02 & 0(0)-\textgreater{}1($1\nu_{11}$)    \\
		& 20492.7 & 532.3    & 2.54E-04 & 0(0)-\textgreater{}1($2\nu_{11}$)       \\ \hline
		1      & 20149.1 & 875.9    & 8.88E-02 & 0(0)-\textgreater{}1($1\nu_{14}$)  \\
		& 19910.4 & 238.7    & 2.93E-04 & 0(0)-\textgreater{}1($1\nu_{8}$, $1\nu_{12}$,$1\nu_{14}$) \\
		& 19883.0 & 266.1    & 1.32E-03 & 0(0)-\textgreater{}1($1\nu_{11}$,$1\nu_{14}$)\\ \hline
		2      & 19274.1 & 875.1    & 6.75E-03 & 0(0)-\textgreater{}1($2\nu_{14}$) \\
		& 19032.9 & 241.1    & 3.44E-05 & 0(0)-\textgreater{}1($1\nu_{9}$, $1\nu_{12}$,$2\nu_{14}$) \\
		& 19007.1 & 267.0    & 9.21E-05 & 0(0)-\textgreater{}1($1\nu_{11}$,$2\nu_{14}$)  \\ \hline
		3      & 18398.2 & 875.9    & 4.07E-04 & 0(0)-\textgreater{}1($3\nu_{14}$)  \\
		& 18159.5 & 238.7    & 1.26E-06 & 0(0)-\textgreater{}1($1\nu_{8}$, $1\nu_{12}$,$3\nu_{14}$) \\
		& 18132.1 & 266.1    & 5.13E-06 & 0(0)-\textgreater{}1($1\nu_{11}$,$3\nu_{14}$)  \\ \hline
		4      & 17522.4 & 875.9    & 2.11E-05 & 0(0)-\textgreater{}1($4\nu_{14}$)  \\
		& 17256.2 & 266.1    & 2.48E-07 & 0(0)-\textgreater{}1($1\nu_{11}$,$4\nu_{14}$)  \\ \bottomrule              
	\end{tabular}
\end{table}
\FloatBarrier

\FloatBarrier
\begin{table}[!htbp]
	\centering
	\caption{Theoretical displacements of geometries $\Delta$ R, and absolute frequency shifts $\Delta \nu$ from the ground to the first excited state.}
	\label{tab:dispalcements}
	\begin{tabular}{llcccccc}
		\toprule
		&     & \multicolumn{2}{c}{$\Delta$ R, \r{A}} & \multicolumn{4}{c}{$\Delta \nu$, \si{\per\cm}} \\ \hline
		&           &  $\mathrm{R_{U-O}}$          &      $\mathrm{R_{U-X}}$             & $\nu_\mathrm{U-X}$                                         & $\nu_\mathrm{b}$ & $\nu_\mathrm{s}$ &$\nu_\mathrm{a}$      \\ \hline
		\ce{[UO2Cl4]^{2-}} & gas phase & 0.031     & 0.007           & -4    & -4    & -80    & -115    \\
		& dodecane  & 0.031     & 0.007           & 1    & -1    & -78    & -106    \\
		& acetone    & 0.03      & 0.009           & -1    & -2    & -81    & -101    \\
		\ce{[UO2Br4]^{2-}}& gas phase & 0.028     & 0.012           & 7    & 2    & -31    & -53     \\
		\ce{[Bu4N]2[UO2Cl4]}& gas phase & 0.026     & 0.003-0.027     &- 4    & -5    & -70    & -100    \\
		\ce{[{A336}]2[UO2Cl4]}& gas phase & 0.029     & 0.005-0.021       & -5    & -4    & -74    & -103   \\ \bottomrule
	\end{tabular}
\end{table}
\FloatBarrier

\clearpage
\section{Supplemental figures}
	\FloatBarrier
	\begin{figure}[!htbp]
		\centering
		\includegraphics[width=.99\textwidth]{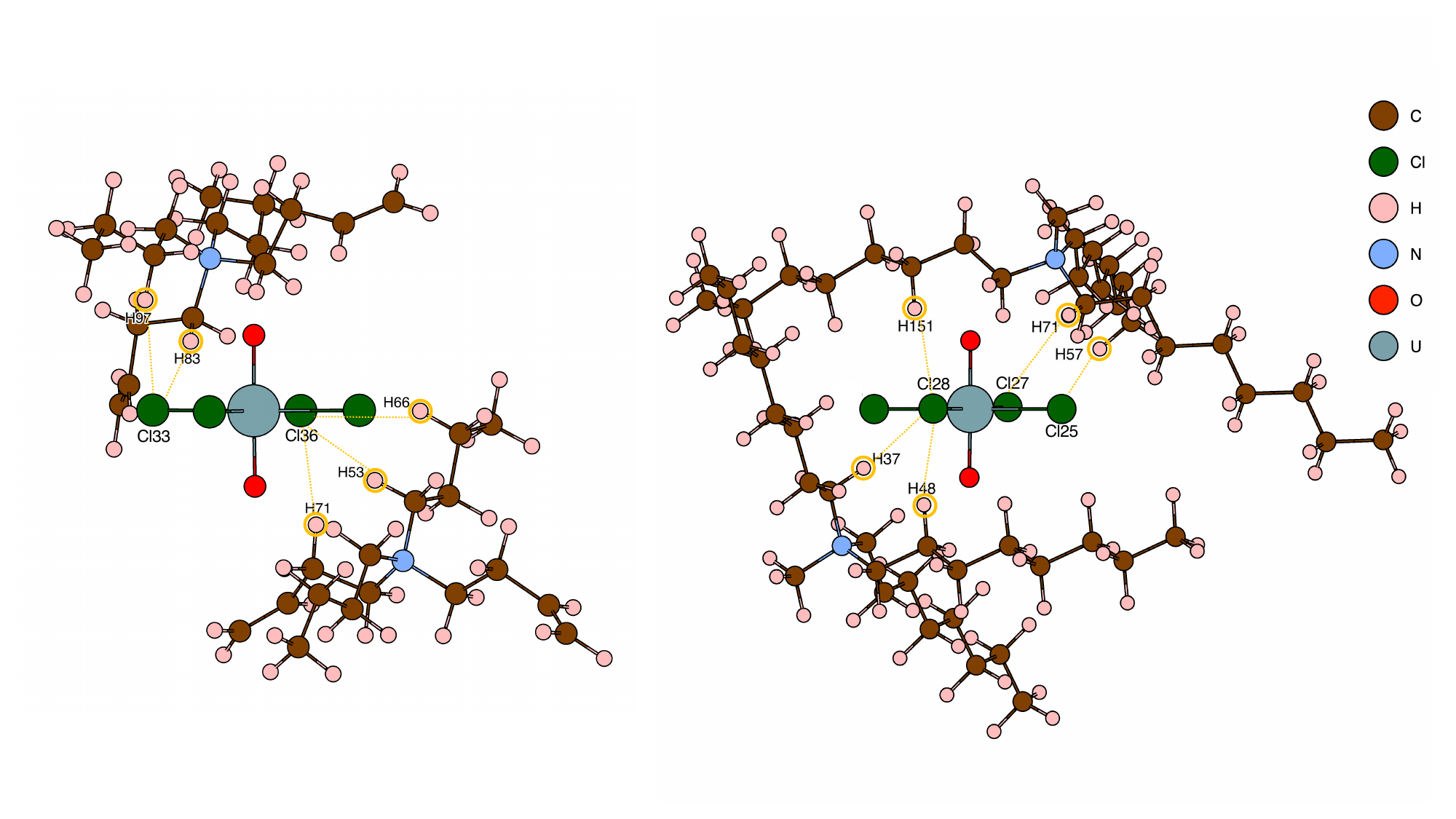}
		\caption{Illustrations of hydrogen bondings between the first and second coordination spheres of uranyl in \ce{[R4N]2[UO2Cl_4]} complexes. Left panel is representing interactions in  \ce{[Bu4N]2[UO2Cl4]} and right in \ce{[{A336}]2[UO2Cl4]}.}
		\label{fig:hyd-bond-bu4n}
	\end{figure}
	\FloatBarrier

\FloatBarrier
\begin{figure}[!htbp]
	\centering
	\includegraphics[width=.8\textwidth]{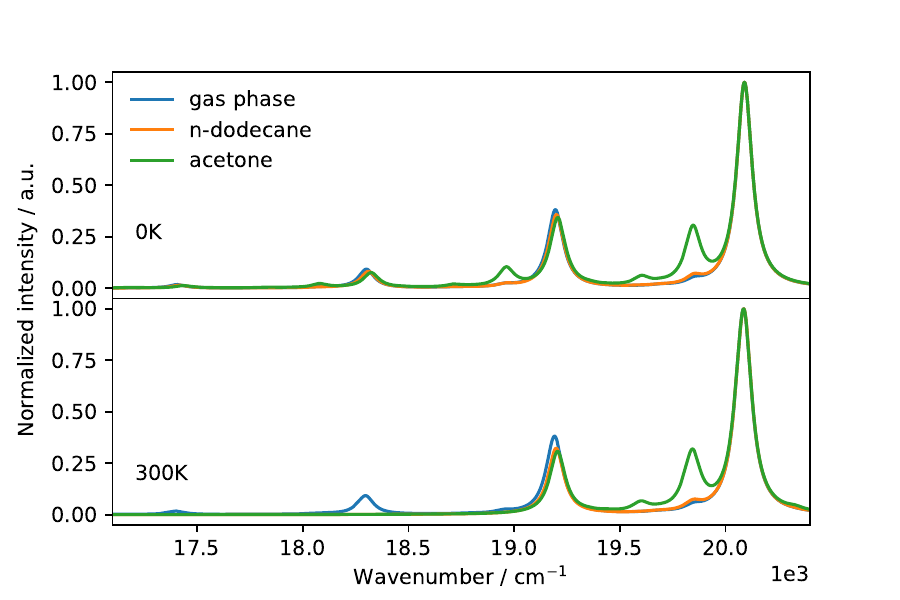}
	\caption{Influence of solvent effects on the theoretical spectrum of \ce{[UO2Cl4]^{2-}} at \SI{0}{\kelvin} (upper panel) and \SI{300}{\kelvin} (lower panel). The spectral shapes were obtained by a Lorentzian convolution; the assignments are provided in Tables \ref{tab:uo2cl42-gas-g16-ass}, \ref{tab:uo2cl42-dod-ass} and \ref{tab:uo2cl42-acetone-ass} for the \ce{[UO2Cl4]^{2-}} in gas-phase, $n$-dodecane and acetone, respectively.}
	\label{fig:solvent-effect-spectra-cl}
\end{figure}
\FloatBarrier

\FloatBarrier
\begin{figure}[!htbp]
	\centering
	\includegraphics[width=.8\textwidth]{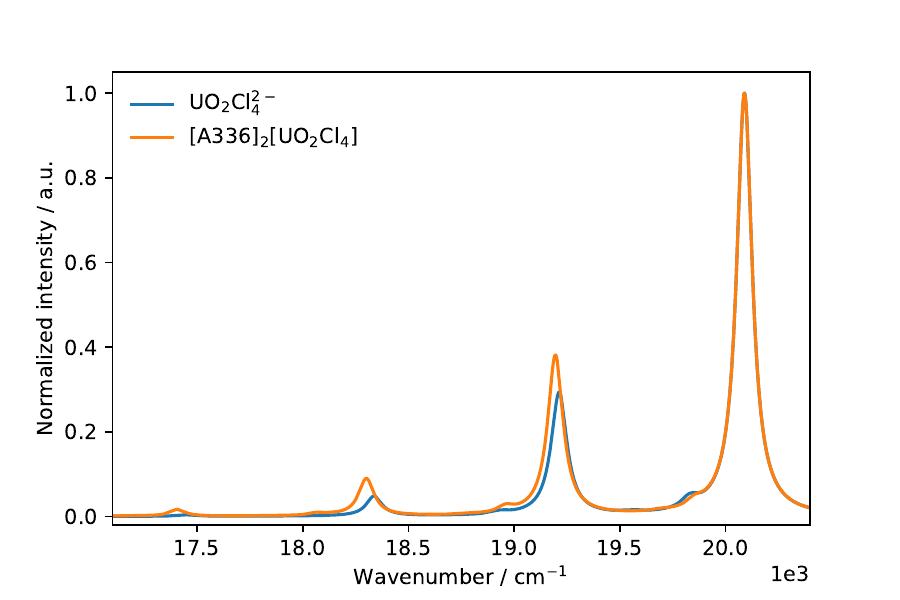}
	\caption{Effect of a \ce{[{A336}]^+} counter ion on the theoretical \ce{[UO2Cl4]^{2-}} spectrum in gas-phase at \SI{300}{\kelvin}. The spectral shapes were obtained by a Lorentzian convolution; the assignments are provided in Tables \ref{tab:uo2cl42-gas-g16-ass} and \ref{tab:uo2cl42-a336-ass} for the \ce{[UO2Cl4]^{2-}} and  \ce{[{A336}]2[UO2Cl4]}, respectively.}
	\label{fig:Cl-A336}
\end{figure}
\FloatBarrier

\FloatBarrier
\begin{figure}[!htbp]
	\centering
	\includegraphics[width=.8\textwidth]{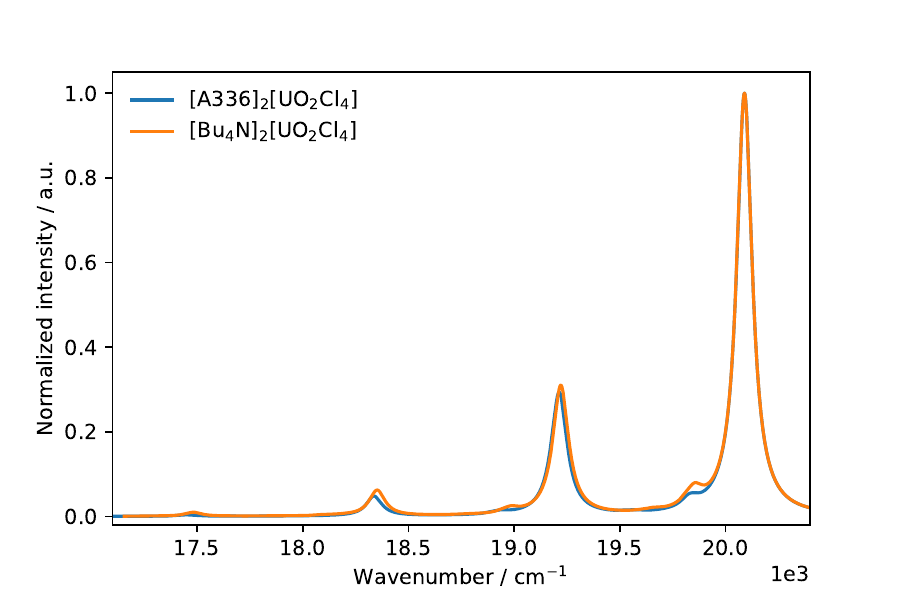}
	\caption{Comparison of the theoretical spectra of \ce{[Bu4N]2[UO2Cl4]} and \ce{[{A336}]2[UO2Cl4]} complexes in a gas-phase at \SI{300}{\kelvin}. The spectral shapes were obtained by a Lorentzian convolution; the assignments are provided in Tables \ref{tab:uo2cl42-bu4n-ass} and \ref{tab:uo2cl42-a336-ass} for the \ce{[Bu4N]2[UO2Cl4]} and \ce{[{A336}]2[UO2Cl4]}, respectively.}
	\label{fig:counter ion}
\end{figure}
\FloatBarrier

	\FloatBarrier
	\begin{table}[!htbp]
		\centering
		\caption{Cartesian coordinates (in \r{A}) of the \ce{[UO2X4]^{2-}} complexes in different media at their ground and first excited state minima obtained by R-ECP DFT/PBE0 and TD-DFT/PBE0 methods, respectively.}
		\label{tab:coordinated-first sphere}
		\begin{tabular}{cccc|ccc}
			\toprule
			&\multicolumn{6}{l}{\ce{[UO2Cl4]^{2-}} gas phase, Turbomole 7.3.1}                                     \\ \hline
			& \multicolumn{3}{c}{Ground state}      &   \multicolumn{3}{c}{Excited state}    \\
			Cl &   1.919262  & 1.919262   & 0          &   1.9246369  & 1.9246369  & 0          \\
			Cl & -1.919262   & -1.919262  & 0          &  -1.9246369 & -1.9246369 & 0          \\
			Cl & 1.919262    & -1.919262  & 0          &   1.9246369  & -1.9246369 & 0          \\
			Cl & -1.919262   & 1.919262   & 0          &   -1.9246369 & 1.9246369  & 0          \\
			O  & 0           & 0          & -1.7581059 &   0          & 0          & -1.7888649 \\
			O  & 0           & 0          & 1.7581059  &   0          & 0          & 1.7888649  \\
			U  & 0           & 0          & 0          &  & 0           0          & 0          \\ \hline
			&\multicolumn{6}{l}{\ce{[UO2Cl4]^{2-}} $n$-dodecane, G16}                                            \\ \hline
			Cl & 0           & 2.698696   & 0          &   0          & 2.706546   & 0          \\
			Cl & 0           & -2.698696  & 0          &   0          & -2.706546  & 0          \\
			Cl & -2.698696   & 0          & 0          &   -2.706546  & 0          & 0          \\
			Cl & 2.698696    & 0          & 0          &   2.706546   & 0          & 0          \\
			O  & 0           & 0          & 1.759301   &   0          & 0          & 1.789492   \\
			O  & 0           & 0          & -1.759301  &   0          & 0          & -1.789492  \\
			U  & 0           & 0          & 0          &   0          & 0          & 0          \\ \hline
			&\multicolumn{6}{l}{\ce{[UO2Cl4]^{2-}} acetone, G16}                                             \\ \hline
			Cl & 0           & 2.683942   & 0          &   0          & 2.691882   & 0          \\
			Cl & 0           & -2.683942  & 0          &   0          & -2.691882  & 0          \\
			Cl & -2.683942   & 0          & 0          &   -2.691882  & 0          & 0          \\
			Cl & 2.683942    & 0          & 0          &  2.691882   & 0          & 0          \\
			O  & 0           & 0          & 1.760965   &  0          & 0          & 1.790628   \\
			O  & 0           & 0          & -1.760965  &   0          & 0          & -1.790628  \\
			U  & 0           & 0          & 0          &   0          & 0          & 0          \\ \hline
			&\multicolumn{6}{l}{\ce{[UO2Br4]^{2-}} gas phase, Turbomole 7.3.1}                                     \\ \hline
			Br & -2.03856600 & -2.0385665 & 0.0000001  &   -2.0466917 & -2.0466919 & -0.0001551 \\
			Br & 2.03856650  & 2.0385659  & 0.0000001  &   2.0466764  & 2.0466945  & -0.0001593 \\
			Br & -2.03856600 & 2.0385658  & 0          &   -2.0467227 & 2.0466596  & 0.0001346  \\
			Br & 2.03856600  & -2.0385662 & 0.0000001  & 2.0466703  & -2.0467211 & 0.0001357  \\
			O  & -0.00000030 & 0.0000005  & -1.7503044 &   0.0000395  & 0.0000376  & -1.7775715 \\
			O  & -0.00000030 & 0.0000005  & 1.7503042  &   0.0000372  & 0.0000294  & 1.7776021  \\
			U  & 0.00000010  & -0.0000001 & -0.0000001 &   -0.0000091 & -0.0000081 & 0.0000135 \\ 
			\bottomrule
		\end{tabular}
	\end{table}
\FloatBarrier

\section{Cartesian coordinates in {\AA} of all the complexes}
\FloatBarrier
	\begin{longtable}{cccc|ccc}
	\caption{Cartesian coordinates (in \r{A}) of the \ce{[Bu4N]2[UO2Cl4]} in gas phase at its ground and first excited state minima obtained in Turbomole 7.3.1 by R-ECP DFT/PBE0 and TD-DFT/PBE0 methods respectively.} \\
	\toprule
	& \multicolumn{3}{c}{Ground state}               & \multicolumn{3}{c}{Excited state}               \\
	\hline
	\endfirsthead
	\multicolumn{7}{c}%
	{\tablename\ \thetable\ -- \textit{Continued from previous page}} \\
	\hline
	& \multicolumn{3}{c}{Ground state}               & \multicolumn{3}{c}{Excited state}               \\
	\hline
	\endhead
	\hline \multicolumn{7}{r}{\textit{Continued on next page}} \\
	\endfoot
	\bottomrule
	\endlastfoot
	C  & -1.7231178 & 2.6706283  & -2.6216748 & -1.6793731 & 2.6734267  & -2.6465738 \\
	C  & -2.0317492 & 1.2075078  & -2.8936115 & -1.9941509 & 1.2097959  & -2.9084194 \\
	C  & -3.3758543 & 1.0920197  & -3.5415702 & -3.3303708 & 1.0965456  & -3.5728047 \\
	C  & -4.4133376 & 0.4600839  & -3.0135895 & -4.3783182 & 0.4748322  & -3.0534051 \\
	C  & -0.3096963 & 4.4588521  & -1.8591334 & -0.2615627 & 4.4594906  & -1.8864919 \\
	C  & -0.1792549 & 2.2495675  & -0.7223308 & -0.1678411 & 2.265401   & -0.7187768 \\
	C  & 0.7403436  & 2.4820091  & -2.943684  & 0.7867802  & 2.4586391  & -2.9290301 \\
	N  & -0.3714943 & 2.9649229  & -2.0406626 & -0.3341219 & 2.9638944  & -2.0490737 \\
	C  & 8.2198311  & 0.6994753  & -3.063383  & 8.1677943  & 0.7167897  & -3.0809801 \\
	C  & 7.1078312  & -0.2278135 & -3.4509097 & 7.0503487  & -0.2071862 & -3.460733  \\
	C  & 6.9992163  & -0.8016964 & -4.6396134 & 6.925001   & -0.7710549 & -4.6525771 \\
	C  & 8.9381358  & 4.0843228  & -4.4134742 & 8.84534    & 4.1199889  & -4.4064407 \\
	C  & 10.0010406 & 4.3693995  & -5.4735881 & 9.8818182  & 4.4204104  & -5.4882432 \\
	C  & 0.7639403  & 3.0479043  & -4.346386  & 0.8343909  & 3.0027985  & -4.3396918 \\
	C  & 1.8951095  & 2.3959104  & -5.1365636 & 1.9678381  & 2.3254385  & -5.1049056 \\
	C  & 2.0237276  & 2.9657649  & -6.5378999 & 2.1229064  & 2.8737564  & -6.5121335 \\
	C  & 0.9646197  & 5.0129345  & -1.2460252 & 1.0009632  & 5.0091596  & -1.2455701 \\
	C  & 0.8803024  & 6.5064323  & -1.2197909 & 0.939701   & 6.5038335  & -1.2584027 \\
	C  & 0.9391448  & 7.2434357  & -0.120433  & 0.9678758  & 7.2674491  & -0.1761263 \\
	C  & -1.173654  & 2.5638109  & 0.3729298  & -1.1765075 & 2.6033288  & 0.3562616  \\
	C  & -0.8081203 & 1.7683579  & 1.624225   & -0.8394835 & 1.822709   & 1.6246737  \\
	C  & -1.7407864 & 2.0567479  & 2.7867969  & -1.7907905 & 2.1336595  & 2.7661672  \\
	C  & 7.7399134  & 4.2181381  & -1.6440369 & 7.7130724  & 4.2168068  & -1.6083406 \\
	C  & 8.4885346  & 5.4138149  & -1.0823067 & 8.4722942  & 5.4061695  & -1.0476814 \\
	C  & 7.5036746  & 6.4986065  & -0.7691173 & 7.4928854  & 6.4846622  & -0.6976385 \\
	C  & 7.5741331  & 7.7276482  & -1.2582746 & 7.551544   & 7.7209691  & -1.1697449 \\
	C  & 7.6175083  & 2.0055634  & -2.5780823 & 7.5738296  & 2.0153299  & -2.5661762 \\
	C  & 9.5478961  & 3.4588312  & -3.1755507 & 9.486048   & 3.4809697  & -3.1912622 \\
	C  & 9.4345973  & 2.524316   & -0.9564187 & 9.4275313  & 2.518378   & -0.9822147 \\
	C  & 8.6866428  & 1.9672626  & 0.2325169  & 8.7091061  & 1.9397153  & 0.2146703  \\
	C  & 9.65996    & 1.6011049  & 1.3469478  & 9.7073695  & 1.5752569  & 1.3073019  \\
	C  & 8.9400933  & 1.0211657  & 2.5532241  & 9.0182853  & 0.9690792  & 2.5186037  \\
	Cl & 5.6570547  & 3.8447947  & 1.2252483  & 5.7037915  & 3.7874771  & 1.2948984  \\
	Cl & 2.6093874  & 0.1291399  & -1.1318178 & 2.5725988  & 0.1000307  & -1.0465703 \\
	Cl & 3.9992506  & 0.5892005  & 2.413443   & 3.9935588  & 0.5548479  & 2.4965168  \\
	Cl & 4.2352165  & 3.317432   & -2.1815109 & 4.2385258  & 3.2897317  & -2.1284497 \\
	N  & 8.5911181  & 3.0516001  & -2.0914457 & 8.5562295  & 3.0576174  & -2.0899549 \\
	O  & 2.6579787  & 2.8136216  & 0.6054672  & 2.6563163  & 2.8170192  & 0.6914103  \\
	O  & 5.5781868  & 1.0410046  & -0.3099148 & 5.5865699  & 0.970901   & -0.2560554 \\
	U  & 4.1153844  & 1.9065225  & 0.1791804  & 4.1164657  & 1.8664524  & 0.2561967  \\
	C  & 11.0137114 & 5.434966   & -5.0831033 & 10.9007723 & 5.4842478  & -5.1095114 \\
	H  & -2.4559006 & 3.0785152  & -1.9256472 & -2.419443  & 3.0935144  & -1.9657071 \\
	H  & -1.7982081 & 3.2485531  & -3.5434355 & -1.7366446 & 3.2425793  & -3.5750709 \\
	H  & -1.2748837 & 0.7780204  & -3.5571044 & -1.2315038 & 0.7691541  & -3.5577871 \\
	H  & -2.0215924 & 0.6273467  & -1.968997  & -1.9995663 & 0.6384053  & -1.9782886 \\
	H  & -3.4833213 & 1.5690257  & -4.5143349 & -3.4222629 & 1.5658076  & -4.5509215 \\
	H  & -4.3470267 & -0.033883  & -2.0491824 & -4.3275788 & -0.0114333 & -2.0841513 \\
	H  & -5.3636916 & 0.4031782  & -3.5307445 & -5.3220169 & 0.418791   & -3.5826978 \\
	H  & -1.1694865 & 4.7270943  & -1.2442715 & -1.1354452 & 4.7457184  & -1.3001097 \\
	H  & -0.4747673 & 4.889107   & -2.8484588 & -0.3939071 & 4.877186   & -2.8859806 \\
	H  & 0.8313591  & 2.485545   & -0.3933224 & 0.840782   & 2.4954168  & -0.3797986 \\
	H  & -0.1703213 & 1.1851182  & -0.9538556 & -0.163531  & 1.1975964  & -0.9348487 \\
	H  & 1.6781943  & 2.7042229  & -2.4331307 & 1.7199389  & 2.6815295  & -2.409592  \\
	H  & 0.6705955  & 1.3951138  & -2.9564868 & 0.7084254  & 1.3722497  & -2.9260715 \\
	H  & 8.8870508  & 0.8395134  & -3.9189523 & 8.8181177  & 0.8698745  & -3.9472294 \\
	H  & 8.8005869  & 0.2315785  & -2.2604406 & 8.7652978  & 0.2387174  & -2.2964379 \\
	H  & 6.3554238  & -0.4110288 & -2.6867537 & 6.3082732  & -0.3975701 & -2.6882462 \\
	H  & 7.7295906  & -0.6321114 & -5.4257046 & 7.6439182  & -0.5943258 & -5.4475929 \\
	H  & 6.1758188  & -1.4679127 & -4.8672707 & 6.0987361  & -1.4358077 & -4.8740409 \\
	H  & 8.1861272  & 3.4175562  & -4.8417619 & 8.0844216  & 3.4567764  & -4.8243736 \\
	H  & 8.4286978  & 5.0183785  & -4.1607759 & 8.3401671  & 5.0496251  & -4.130119  \\
	H  & 10.5203372 & 3.4386731  & -5.7291638 & 10.3972858 & 3.4942487  & -5.7671456 \\
	H  & 9.4852783  & 4.6841782  & -6.3848823 & 9.3435881  & 4.7443912  & -6.3831726 \\
	H  & -0.1819632 & 2.8781471  & -4.8734487 & -0.1066389 & 2.8364979  & -4.8766122 \\
	H  & 0.9321954  & 4.1295991  & -4.3262175 & 1.0160613  & 4.0824947  & -4.3337166 \\
	H  & 2.8304257  & 2.5281695  & -4.5847193 & 2.8966736  & 2.4542972  & -4.5414708 \\
	H  & 1.7209957  & 1.3158409  & -5.1880289 & 1.7806363  & 1.2470119  & -5.1436291 \\
	H  & 2.2450226  & 4.0363524  & -6.5097535 & 2.3576975  & 3.9417306  & -6.4957196 \\
	H  & 1.1029825  & 2.8320793  & -7.1138735 & 1.2084956  & 2.7438894  & -7.0990188 \\
	H  & 2.8306813  & 2.4769339  & -7.0874904 & 2.9308802  & 2.3669528  & -7.0437125 \\
	H  & 1.8452075  & 4.7093883  & -1.8236081 & 1.896008   & 4.677173   & -1.7843465 \\
	H  & 1.1193957  & 4.6333552  & -0.2350723 & 1.1145126  & 4.6539952  & -0.2202624 \\
	H  & 0.7760567  & 7.0020013  & -2.1843963 & 0.8806151  & 6.9763318  & -2.2382199 \\
	H  & 1.0627443  & 6.7908448  & 0.8582337  & 1.0462638  & 6.838126   & 0.8175987  \\
	H  & 0.8916486  & 8.3253259  & -0.1622608 & 0.9394954  & 8.3485261  & -0.2462526 \\
	H  & -2.1998613 & 2.3154108  & 0.0783695  & -2.1995984 & 2.3587715  & 0.0481891  \\
	H  & -1.1602673 & 3.6301845  & 0.6214414  & -1.1592043 & 3.6731832  & 0.5893362  \\
	H  & 0.2246834  & 1.9950809  & 1.9040768  & 0.1897737  & 2.0454302  & 1.9203767  \\
	H  & -0.8252225 & 0.6989652  & 1.3896435  & -0.8598857 & 0.7504031  & 1.4042654  \\
	H  & -1.7098422 & 3.1134609  & 3.0666007  & -1.7570608 & 3.1937903  & 3.0324141  \\
	H  & -1.4578685 & 1.4754306  & 3.6661139  & -1.5285628 & 1.5620758  & 3.6581993  \\
	H  & -2.7777888 & 1.8077791  & 2.54197    & -2.824966  & 1.8890225  & 2.505678   \\
	H  & 7.0416344  & 3.8371666  & -0.8966282 & 7.0314797  & 3.8249679  & -0.8510347 \\
	H  & 7.1386468  & 4.5036323  & -2.5057393 & 7.0918112  & 4.5117038  & -2.4524489 \\
	H  & 9.2412281  & 5.799676   & -1.7765467 & 9.2087551  & 5.8030053  & -1.7531317 \\
	H  & 8.9983067  & 5.1322407  & -0.1544231 & 9.0029946  & 5.1135421  & -0.1350372 \\
	H  & 6.6948592  & 6.215431   & -0.0989246 & 6.6991158  & 6.1899392  & -0.0145572 \\
	H  & 8.3714955  & 8.0315552  & -1.9308543 & 8.3339235  & 8.0359822  & -1.8547064 \\
	H  & 6.8389032  & 8.4794786  & -0.9972636 & 6.8212509  & 8.4677866  & -0.8818384 \\
	H  & 6.9340373  & 1.8155002  & -1.7513022 & 6.9086759  & 1.8163625  & -1.7267325 \\
	H  & 7.0229452  & 2.4744519  & -3.3603095 & 6.9587697  & 2.4920396  & -3.3274802 \\
	H  & 10.2577446 & 4.1412972  & -2.7092372 & 10.2048478 & 4.1596966  & -2.7332091 \\
	H  & 10.1018386 & 2.5586908  & -3.4467517 & 10.0359903 & 2.5860505  & -3.4868352 \\
	H  & 10.0744458 & 3.3546276  & -0.6526366 & 10.069342  & 3.3476496  & -0.679945  \\
	H  & 10.0864893 & 1.7656666  & -1.3945915 & 10.0737623 & 1.7699849  & -1.4457339 \\
	H  & 8.1107541  & 1.0823836  & -0.0477573 & 8.1407886  & 1.0493279  & -0.0645889 \\
	H  & 7.9567762  & 2.6857104  & 0.6174913  & 7.9761006  & 2.6434691  & 0.6202086  \\
	H  & 10.2226809 & 2.4932853  & 1.6477165  & 10.2603168 & 2.4730125  & 1.6093349  \\
	H  & 10.4011149 & 0.8825933  & 0.9745528  & 10.453713  & 0.873802   & 0.9133679  \\
	H  & 8.2046395  & 1.7265971  & 2.9456944  & 8.2763823  & 1.6565146  & 2.9304332  \\
	H  & 9.6434408  & 0.778559   & 3.3526434  & 9.7389617  & 0.7311515  & 3.3038452  \\
	H  & 8.3999497  & 0.109224   & 2.2890534  & 8.491638   & 0.0495992  & 2.2530673  \\
	H  & 11.634522  & 5.131161   & -4.2367324 & 11.5420253 & 5.1725295  & -4.2814456 \\
	H  & 10.5154962 & 6.3702383  & -4.8128049 & 10.4066819 & 6.4149568  & -4.8168367 \\
	H  & 11.6894051 & 5.6503302  & -5.9130371 & 11.5562527 & 5.7109501  & -5.9525399 \\
\end{longtable}
	\FloatBarrier

\FloatBarrier
	\begin{longtable}{cccc|ccc}
	\caption{Cartesian coordinates (in \r{A}) of the \ce{[{A336}]2[UO2Cl4]} in gas phase at its ground and first excited state minima obtained in Turbomole 7.3.1. by R-ECP DFT/PBE0 and TD-DFT/PBE0 methods respectively.}\\
	\toprule
	& \multicolumn{3}{c}{Ground state}               & \multicolumn{3}{c}{Excited state}               \\
	\hline
	\endfirsthead
	\multicolumn{7}{c}%
	{\tablename\ \thetable\ -- \textit{Continued from previous page}} \\
	\hline
	& \multicolumn{3}{c}{Ground state}               & \multicolumn{3}{c}{Excited state}               \\
	\hline
	\endhead
	\hline \multicolumn{7}{r}{\textit{Continued on next page}} \\
	\endfoot
	\bottomrule
	\endlastfoot
	C    & -1.6461843 & 1.2032269  & -2.7898103  & -1.633078  & 1.1910779  & -2.8057852  \\
	C    & -0.5220577 & 3.2945601  & -2.294429   & -0.532041  & 3.2905882  & -2.295121   \\
	C    & -0.0115532 & 1.2329116  & -0.9622376  & -0.0219215 & 1.2369558  & -0.9570139  \\
	C    & 0.7562838  & 1.3638802  & -3.2560934  & 0.7708408  & 1.3704842  & -3.2431944  \\
	N    & -0.3658949 & 1.7925459  & -2.3275443  & -0.3642496 & 1.7906856  & -2.3259703  \\
	C    & 8.1975976  & 1.7276116  & -3.1163927  & 8.164239   & 1.7028967  & -3.134453   \\
	C    & 7.057662   & 1.1507649  & -3.9665424  & 7.0107299  & 1.1265578  & -3.9654652  \\
	C    & 0.7077407  & 1.8813497  & -4.6808654  & 0.7500881  & 1.9200359  & -4.6562278  \\
	C    & 1.8537649  & 1.2580073  & -5.4787245  & 1.8757458  & 1.2730443  & -5.4628106  \\
	C    & 1.9286199  & 1.7585894  & -6.9158336  & 1.9844609  & 1.814585   & -6.8823089  \\
	C    & 0.6900871  & 4.0531035  & -1.797232   & 0.654431   & 4.0591905  & -1.7525596  \\
	C    & 0.495246   & 5.5655171  & -1.8656286  & 0.4762728  & 5.5670984  & -1.9064005  \\
	C    & -0.9271163 & 1.5988199  & 0.1906763   & -0.979795  & 1.5620885  & 0.1731472   \\
	C    & -0.4031856 & 0.9528474  & 1.474797    & -0.4617339 & 0.9389285  & 1.4703199   \\
	C    & -1.223735  & 1.3168107  & 2.7064365   & -1.3267301 & 1.269012   & 2.6805805   \\
	C    & 8.2115209  & 5.2420106  & -1.590463   & 8.217084   & 5.2184853  & -1.6148665  \\
	C    & 9.0401054  & 6.2481231  & -0.8140585  & 9.063911   & 6.2190133  & -0.851928   \\
	C    & 8.1481801  & 7.4186072  & -0.3968068  & 8.1873798  & 7.3976505  & -0.4271156  \\
	C    & 7.8496605  & 3.1628947  & -2.7690856  & 7.8235677  & 3.1374496  & -2.7781333  \\
	C    & 9.9606169  & 4.360842   & -3.0775044  & 9.9325979  & 4.3263012  & -3.1325281  \\
	C    & 9.5942091  & 3.2511344  & -0.9563462  & 9.6041048  & 3.2244248  & -1.0023766  \\
	C    & 8.7007022  & 2.869779   & 0.204574    & 8.7341427  & 2.8461309  & 0.1769905   \\
	C    & 9.4294255  & 1.9580388  & 1.1899717   & 9.4918775  & 1.956401   & 1.1600231   \\
	C    & 8.5778865  & 1.6647567  & 2.4210161   & 8.6678887  & 1.666431   & 2.4100975   \\
	Cl   & 5.3884893  & 4.2984281  & 0.3064566   & 5.4510299  & 4.2706622  & 0.3366167   \\
	Cl   & 3.2420551  & -0.2944559 & -1.3911706  & 3.1912206  & -0.3097982 & -1.3195414  \\
	Cl   & 4.7608904  & 0.8807181  & 1.9952177   & 4.8012604  & 0.8545479  & 2.0403593   \\
	Cl   & 3.985672   & 3.0256857  & -2.8432433  & 3.9528578  & 3.0244947  & -2.802491   \\
	N    & 8.9204324  & 3.9995534  & -2.0835293  & 8.9107053  & 3.9715808  & -2.1176084  \\
	O    & 2.7388955  & 2.4540567  & 0.1199908   & 2.7296024  & 2.4763278  & 0.1993843   \\
	O    & 5.9502491  & 1.3823912  & -0.8952207  & 5.9526931  & 1.3344897  & -0.8900281  \\
	U    & 4.3485499  & 1.9091212  & -0.3673548  & 4.3466912  & 1.8905967  & -0.3161897  \\
	H    & -1.4046432 & 3.4946698  & -1.6658059  & -1.4383236 & 3.4841597  & -1.6985912  \\
	H    & -0.7817986 & 3.5928913  & -3.3225549  & -0.7564003 & 3.5894676  & -3.3313615  \\
	H    & 1.012849   & 1.5669308  & -0.7437302  & 0.9855016  & 1.605448   & -0.7147684  \\
	H    & 0.0389693  & 0.1413441  & -1.096228   & 0.0756246  & 0.1487302  & -1.0937655  \\
	H    & 1.7025041  & 1.6699581  & -2.784895   & 1.713517   & 1.6572178  & -2.75183    \\
	H    & 0.7455572  & 0.2633428  & -3.2346128  & 0.7482958  & 0.269755   & -3.2454746  \\
	H    & 9.1608184  & 1.6398624  & -3.6540545  & 9.1184122  & 1.6155065  & -3.6882649  \\
	H    & 8.276449   & 1.1382993  & -2.1884052  & 8.2617595  & 1.1112358  & -2.2091603  \\
	H    & -0.2528237 & 1.6494442  & -5.1786956  & -0.2157256 & 1.7377552  & -5.1651685  \\
	H    & 0.8310311  & 2.9785813  & -4.6943621  & 0.9167018  & 3.0116095  & -4.6424251  \\
	H    & 2.8019028  & 1.4814217  & -4.957075   & 2.8268654  & 1.4427365  & -4.9265565  \\
	H    & 1.7477237  & 0.1563279  & -5.4765198  & 1.7242635  & 0.1769694  & -5.4952025  \\
	H    & 2.0461717  & 2.8595914  & -6.9120747  & 2.1495802  & 2.9087937  & -6.8417109  \\
	H    & 0.9709829  & 1.5539303  & -7.4348059  & 1.0227516  & 1.6691845  & -7.4141519  \\
	C    & 3.0737849  & 1.1344469  & -7.7059526  & 3.1073145  & 1.1669667  & -7.6842936  \\
	H    & 1.588215   & 3.7853979  & -2.3811357  & 1.5883639  & 3.7600546  & -2.2613403  \\
	H    & 0.9247961  & 3.769382   & -0.7589082  & 0.8126561  & 3.822178   & -0.68769    \\
	H    & -1.9691455 & 1.2705236  & 0.0164803   & -2.001234  & 1.1874134  & -0.0269822  \\
	H    & -0.9495403 & 2.6933871  & 0.3356569   & -1.056413  & 2.6548985  & 0.3161883   \\
	H    & 0.6484942  & 1.2578452  & 1.624974    & 0.5735104  & 1.2844171  & 1.6462799   \\
	H    & -0.3878051 & -0.1466636 & 1.3526675   & -0.3978414 & -0.1589915 & 1.349348    \\
	H    & -1.2285351 & 2.4174772  & 2.8286919   & -1.3800557 & 2.368776   & 2.801683    \\
	C    & -0.7000726 & 0.6728983  & 3.9854928   & -0.8100195 & 0.64662    & 3.972517    \\
	H    & -2.2806273 & 1.0212465  & 2.5518233   & -2.3666843 & 0.9309914  & 2.4992819   \\
	H    & 7.3602861  & 4.8929891  & -0.9772772  & 7.3722848  & 4.8753488  & -0.9888143  \\
	H    & 7.784598   & 5.7082526  & -2.4937803  & 7.7798279  & 5.687144   & -2.5121589  \\
	H    & 9.8914018  & 6.6306152  & -1.4077287  & 9.9115202  & 6.5925165  & -1.4568101  \\
	H    & 9.458978   & 5.7836087  & 0.0958255   & 9.4896742  & 5.7532647  & 0.0543027   \\
	H    & 6.9730742  & 3.1678889  & -2.1054602  & 6.9617209  & 3.1426009  & -2.0946294  \\
	H    & 7.5773403  & 3.7324783  & -3.6724679  & 7.5304103  & 3.7079881  & -3.6745387  \\
	H    & 10.4257158 & 3.8925993  & -0.622177   & 10.4426746 & 3.8655498  & -0.6849707  \\
	H    & 10.0391224 & 2.3562154  & -1.4185107  & 10.039784  & 2.32803    & -1.4709907  \\
	H    & 7.7968588  & 2.3505228  & -0.1526226  & 7.8287635  & 2.3145072  & -0.1580092  \\
	H    & 8.3423506  & 3.7693938  & 0.7321074   & 8.371927   & 3.7470585  & 0.6998276   \\
	H    & 10.3930872 & 2.412322   & 1.4982492   & 10.454286  & 2.4282738  & 1.4455823   \\
	H    & 9.6845888  & 1.0062478  & 0.6840358   & 9.7530335  & 1.0018581  & 0.6619645   \\
	H    & 8.4108146  & 2.6074578  & 2.9765874   & 8.4883168  & 2.6149005  & 2.9518627   \\
	C    & 9.1797033  & 0.6234488  & 3.3562885   & 9.3095754  & 0.6560275  & 3.3521519   \\
	H    & 7.5745105  & 1.3288947  & 2.1009016   & 7.6666959  & 1.3038002  & 2.1127397   \\
	H    & -1.8902656 & 1.5901533  & -3.7883501  & -1.8540593 & 1.5597719  & -3.8167565  \\
	H    & -2.447729  & 1.4721236  & -2.088683   & -2.4516826 & 1.4706455  & -2.128819   \\
	H    & -1.5395344 & 0.1093463  & -2.8303285  & -1.5236557 & 0.0966638  & -2.8244662  \\
	H    & 9.4976102  & 4.9488504  & -3.883417   & 9.4549004  & 4.9093163  & -3.9337219  \\
	H    & 10.3964441 & 3.4420853  & -3.4938705  & 10.3615574 & 3.4052695  & -3.551459   \\
	H    & 10.7487221 & 4.9522443  & -2.5917938  & 10.7293881 & 4.9208247  & -2.6647543  \\
	H    & -0.7042952 & -0.4280191 & 3.8675313   & -0.7662474 & -0.453641  & 3.8553753   \\
	C    & -1.4954405 & 1.0499705  & 5.2301499   & -1.6511198 & 0.9910132  & 5.1959948   \\
	H    & 0.3602227  & 0.9565859  & 4.1253519   & 0.2342693  & 0.9728038  & 4.1390915   \\
	H    & -2.558063  & 0.7696984  & 5.0870388   & -2.6976975 & 0.6680509  & 5.0261044   \\
	H    & -1.4865368 & 2.1512956  & 5.3496641   & -1.6903862 & 2.0919542  & 5.3145543   \\
	C    & -0.9738292 & 0.4033963  & 6.5092306   & -1.1365904 & 0.3661347  & 6.488451    \\
	H    & -0.9830213 & -0.6967159 & 6.3893882   & -1.0974087 & -0.733508  & 6.3691767   \\
	C    & -1.7685197 & 0.7871262  & 7.7500482   & -1.9777828 & 0.7166822  & 7.70804     \\
	H    & 0.0877397  & 0.6819188  & 6.6482878   & -0.0909875 & 0.6876049  & 6.6545314   \\
	H    & -2.8284367 & 0.4885377  & 7.6560217   & -3.0220271 & 0.3750094  & 7.5872392   \\
	H    & -1.7460945 & 1.8790966  & 7.9178612   & -2.004576  & 1.8087774  & 7.8755857   \\
	H    & -1.3642061 & 0.3023905  & 8.655625    & -1.5775278 & 0.2488083  & 8.6243815   \\
	C    & 3.17242    & 1.6337896  & -9.1428483  & 3.2392894  & 1.7093182  & -9.1024209  \\
	H    & 2.95861    & 0.0329785  & -7.7076248  & 2.9443251  & 0.0720774  & -7.7238015  \\
	H    & 4.0265774  & 1.3350572  & -7.1795698  & 4.063717   & 1.3077747  & -7.1448971  \\
	H    & 2.2179737  & 1.4321839  & -9.6682336  & 2.281857   & 1.5654449  & -9.6415934  \\
	C    & 4.3196435  & 1.0105598  & -9.9313322  & 4.3666896  & 1.0665097  & -9.9032091  \\
	H    & 3.2885062  & 2.7355422  & -9.1401465  & 3.4000453  & 2.8048336  & -9.0616359  \\
	H    & 4.2030578  & -0.0899241 & -9.9325619  & 4.2064682  & -0.0279633 & -9.9413383  \\
	H    & 5.271899   & 1.2112908  & -9.4044095  & 5.3222122  & 1.2114869  & -9.3637366  \\
	C    & 4.4147551  & 1.5139127  & -11.3652394 & 4.4930493  & 1.6120298  & -11.319002  \\
	H    & 3.4900981  & 1.2942647  & -11.9289699 & 3.564844   & 1.4478401  & -11.8959722 \\
	H    & 5.2543813  & 1.0429706  & -11.9054404 & 5.3177593  & 1.1262492  & -11.8692561 \\
	H    & 4.5686073  & 2.6077455  & -11.3964973 & 4.6899982  & 2.6995458  & -11.3134121 \\
	H    & 0.2481085  & 5.8688723  & -2.9025826  & 0.3638107  & 5.8216293  & -2.9792051  \\
	C    & 1.7498638  & 6.3032404  & -1.4059537  & 1.6633694  & 6.3293623  & -1.3256675  \\
	H    & -0.3685886 & 5.8693765  & -1.2410399  & -0.4592259 & 5.8981767  & -1.411689   \\
	H    & 2.5997055  & 5.9897795  & -2.0397849  & 2.5925927  & 5.9529043  & -1.7918052  \\
	H    & 2.0156819  & 5.9683076  & -0.3861392  & 1.7598216  & 6.0851677  & -0.2510504  \\
	C    & 1.621893   & 7.821337   & -1.4266171  & 1.5733997  & 7.840251   & -1.4986194  \\
	H    & 1.3468209  & 8.1576321  & -2.4461799  & 1.5007624  & 8.0848449  & -2.5771766  \\
	C    & 2.8977845  & 8.5332185  & -0.9893175  & 2.7591406  & 8.583014   & -0.8928346  \\
	H    & 0.7860105  & 8.132416   & -0.7690079  & 0.635684   & 8.213194   & -1.0400982  \\
	H    & 3.7282081  & 8.2282757  & -1.6554963  & 3.6961867  & 8.1925409  & -1.3353629  \\
	H    & 3.1832192  & 8.1797494  & 0.0201679   & 2.821807   & 8.3476725  & 0.1873177   \\
	C    & 2.7878661  & 10.0541333 & -0.9817578  & 2.7025212  & 10.0953551 & -1.0779816  \\
	H    & 2.4966596  & 10.4043774 & -1.9908138  & 2.6465683  & 10.3285722 & -2.1589357  \\
	C    & 4.0723714  & 10.7513034 & -0.5533671  & 3.8878001  & 10.8256493 & -0.4608366  \\
	H    & 1.9628702  & 10.3561385 & -0.3084341  & 1.7627083  & 10.4815597 & -0.6382872  \\
	H    & 4.9088156  & 10.4991218 & -1.2304482  & 4.8424853  & 10.4837225 & -0.9010325  \\
	H    & 4.3726872  & 10.4469288 & 0.4655326   & 3.9464154  & 10.6449917 & 0.6278372   \\
	H    & 3.9598992  & 11.8496994 & -0.555265   & 3.8194902  & 11.9168783 & -0.6154867  \\
	H    & 9.3290653  & -0.3217476 & 2.7981355   & 9.4714886  & -0.295812  & 2.8086214   \\
	C    & 8.3064351  & 0.3489486  & 4.5760302   & 8.4660173  & 0.384457   & 4.5928318   \\
	H    & 10.1879736 & 0.9461786  & 3.6862418   & 10.316117  & 1.0077915  & 3.6572149   \\
	H    & 7.2905845  & 0.0738511  & 4.234217    & 7.4502521  & 0.0795998  & 4.2762154   \\
	H    & 8.1846111  & 1.2856827  & 5.1546271   & 8.3326027  & 1.3288745  & 5.1563825   \\
	C    & 8.8526336  & -0.7404977 & 5.4923957   & 9.0549707  & -0.6743837 & 5.5181695   \\
	H    & 8.9661239  & -1.677251  & 4.9137407   & 9.1809585  & -1.6188152 & 4.9544781   \\
	C    & 7.9757058  & -1.000942  & 6.7099023   & 8.2075473  & -0.9331157 & 6.756523    \\
	H    & 9.8727278  & -0.4642193 & 5.82345     & 10.0741762 & -0.3675641 & 5.8245524   \\
	H    & 6.9588529  & -1.3095158 & 6.4082246   & 7.1932039  & -1.2728153 & 6.4802512   \\
	H    & 7.8734476  & -0.092412  & 7.3306414   & 8.0935955  & -0.015587  & 7.3620768   \\
	H    & 8.3934291  & -1.7983681 & 7.3494375   & 8.6568835  & -1.7074323 & 7.4031953   \\
	H    & 7.7335395  & 7.9032666  & -1.3015992  & 7.7675358  & 7.8846953  & -1.3284139  \\
	C    & 8.871702   & 8.4612572  & 0.447091    & 8.9290822  & 8.4349345  & 0.4068213   \\
	H    & 7.2800284  & 7.0229137  & 0.1620042   & 7.3218095  & 7.010128   & 0.1416234   \\
	H    & 9.7429557  & 8.8539358  & -0.1133702  & 9.7990888  & 8.8172189  & -0.1630025  \\
	H    & 9.2841725  & 7.9759179  & 1.3527107   & 9.3456572  & 7.9478288  & 1.3098248   \\
	C    & 7.9730432  & 9.6206695  & 0.8631596   & 8.0459906  & 9.6036712  & 0.8282279   \\
	H    & 7.5573021  & 10.1043126 & -0.0422156  & 7.6278218  & 10.0903772 & -0.074816   \\
	C    & 8.6770028  & 10.669575  & 1.7162676   & 8.7688776  & 10.6455829 & 1.6734333   \\
	H    & 7.1017052  & 9.2219058  & 1.4171583   & 7.1750414  & 9.2147353  & 1.3899872   \\
	H    & 9.5487179  & 11.0696472 & 1.1616067   & 9.6415656  & 11.0335952 & 1.1114653   \\
	H    & 9.0912218  & 10.1860095 & 2.6226454   & 9.1843053  & 10.1582954 & 2.5774414   \\
	C    & 7.7732165  & 11.8254896 & 2.1328072   & 7.8833334  & 11.8133195 & 2.0951097   \\
	H    & 7.3575154  & 12.3058829 & 1.2265388   & 7.4678153  & 12.2987855 & 1.191301    \\
	C    & 8.4796916  & 12.8705764 & 2.9854278   & 8.6095237  & 12.8492746 & 2.9419921   \\
	H    & 6.9031796  & 11.4233346 & 2.6858029   & 7.0116265  & 11.4230405 & 2.6541581   \\
	H    & 9.3334811  & 13.3182153 & 2.4452879   & 9.4660716  & 13.2847477 & 2.3959385   \\
	H    & 8.87366    & 12.4270729 & 3.9175801   & 9.0036359  & 12.4012573 & 3.8721348   \\
	H    & 7.7970997  & 13.6904076 & 3.2687092   & 7.9404078  & 13.6789762 & 3.2292191   \\
	H    & 7.2044857  & 1.4401424  & -5.025524   & 7.1337586  & 1.4263597  & -5.0247294  \\
	H    & 6.1087275  & 1.6176728  & -3.647695   & 6.0659397  & 1.5853394  & -3.6227615  \\
	C    & 6.8959417  & -0.3599666 & -3.8483681  & 6.8613929  & -0.3860122 & -3.8574717  \\
	H    & 7.7901407  & -0.8818376 & -4.2444778  & 7.7466163  & -0.898922  & -4.2848566  \\
	C    & 5.6364194  & -0.8599012 & -4.5475215  & 5.5839534  & -0.8858729 & -4.52193    \\
	H    & 6.8233575  & -0.616522  & -2.7755191  & 6.8226859  & -0.654516  & -2.7855294  \\
	H    & 5.6968024  & -0.6421862 & -5.6330947  & 5.6058445  & -0.6500665 & -5.6052756  \\
	H    & 4.7775496  & -0.2849136 & -4.154714   & 4.734378   & -0.3240671 & -4.0911763  \\
	C    & 5.3616749  & -2.3433383 & -4.33576    & 5.3280228  & -2.3745386 & -4.3262149  \\
	H    & 6.1998702  & -2.9447878 & -4.7414065  & 6.1530055  & -2.962998  & -4.776036   \\
	C    & 4.0510556  & -2.8086398 & -4.9614694  & 3.9958091  & -2.8367164 & -4.9065779  \\
	H    & 5.3274603  & -2.5440842 & -3.2482779  & 5.3398204  & -2.5962546 & -3.2420462  \\
	H    & 4.0557129  & -2.5728676 & -6.0439645  & 3.9554334  & -2.5808401 & -5.9838837  \\
	H    & 3.2256478  & -2.2185758 & -4.5197877  & 3.1860449  & -2.2595093 & -4.4209165  \\
	C    & 3.7770649  & -4.2933944 & -4.76372    & 3.7381282  & -4.3263087 & -4.7258403  \\
	H    & 4.5722748  & -4.9122823 & -5.2179134  & 4.5177204  & -4.9322957 & -5.22297    \\
	H    & 2.8177685  & -4.5957644 & -5.220045   & 2.762759   & -4.6261617 & -5.1488016  \\
	H    & 3.7320634  & -4.5482681 & -3.6897659  & 3.7374553  & -4.6016267 & -3.6558008 
\end{longtable}
\FloatBarrier